\documentclass[11pt]{article}

\usepackage[final]{acl}

\usepackage[table]{xcolor}
\usepackage{microtype}
\usepackage{booktabs}
\usepackage{adjustbox}
\usepackage{tabularx}
\usepackage{multirow}
\usepackage{amsmath}
\usepackage{amssymb}
\usepackage{fancyvrb}
\usepackage{fvextra}
\usepackage{pifont}

\definecolor{blockbest}{RGB}{235,242,250}

\usepackage[english]{babel}
\babelfont{rm}{TeXGyreTermesX}

\title{ParaCalib: Semantically Calibrated Paralinguistic Modeling for Depression Detection}

\author{%
  \textbf{Yuxin Li}\textsuperscript{1,\textdagger},
  \textbf{Yifei Li}\textsuperscript{1,\textdagger},
  \textbf{Yi-Wen Chao}\textsuperscript{1},
  \textbf{Xiangyu Zhang}\textsuperscript{2}
  \\
  \textbf{Eng Siong Chng}\textsuperscript{1},
  \textbf{Cuntai Guan}\textsuperscript{1,*}
  \\[0.6em]
  \textsuperscript{1}Nanyang Technological University, Singapore
  \\
  \textsuperscript{2}University of New South Wales, Sydney, Australia
  \\[0.6em]
  \small{\textsuperscript{\textdagger}These authors contributed equally.}
  \\
  \small{\textsuperscript{*}Corresponding author:
  \href{mailto:CTGuan@ntu.edu.sg}{CTGuan@ntu.edu.sg}}
}
\begin{document}

\maketitle
\begin{abstract}
Vocal behavior provides important signals for speech-based depression detection, but its interpretation often depends on what is being said and how it functions in context. However, existing methods typically treat acoustic cues as context-independent markers, making it difficult to distinguish vocal form from its context-dependent communicative function. We propose ParaCalib, a framework that semantically calibrates paralinguistic behavior by interpreting vocal patterns relative to utterance-level semantic context and inferred communicative function and representing them in a comparable state space. Concretely, ParaCalib uses an Audio-Language Model (ALM) to generate contextualized vocal descriptions and an LLM-based Paralinguistic State Extractor (PSE) to map these descriptions into a structured Semantically Calibrated Paralinguistic (SC-Para) representation. ParaCalib achieves the highest mean Macro-F1 among the evaluated methods, reaching 71.9\% on DAIC-WOZ and 90.5\% on MODMA. Our controlled analysis provides direct evidence of semantic calibration at the PSE stage: under a fixed caption-level acoustic description, varying the accompanying semantic context changes the inferred depression-related paralinguistic evidence. Our exploratory analysis further identifies recurring configurations of paralinguistic states associated with depression labels rather than a single uniformly dominant state.
\end{abstract}

\section{Introduction}

\begin{figure}
    \centering
    \includegraphics[width=1\linewidth]{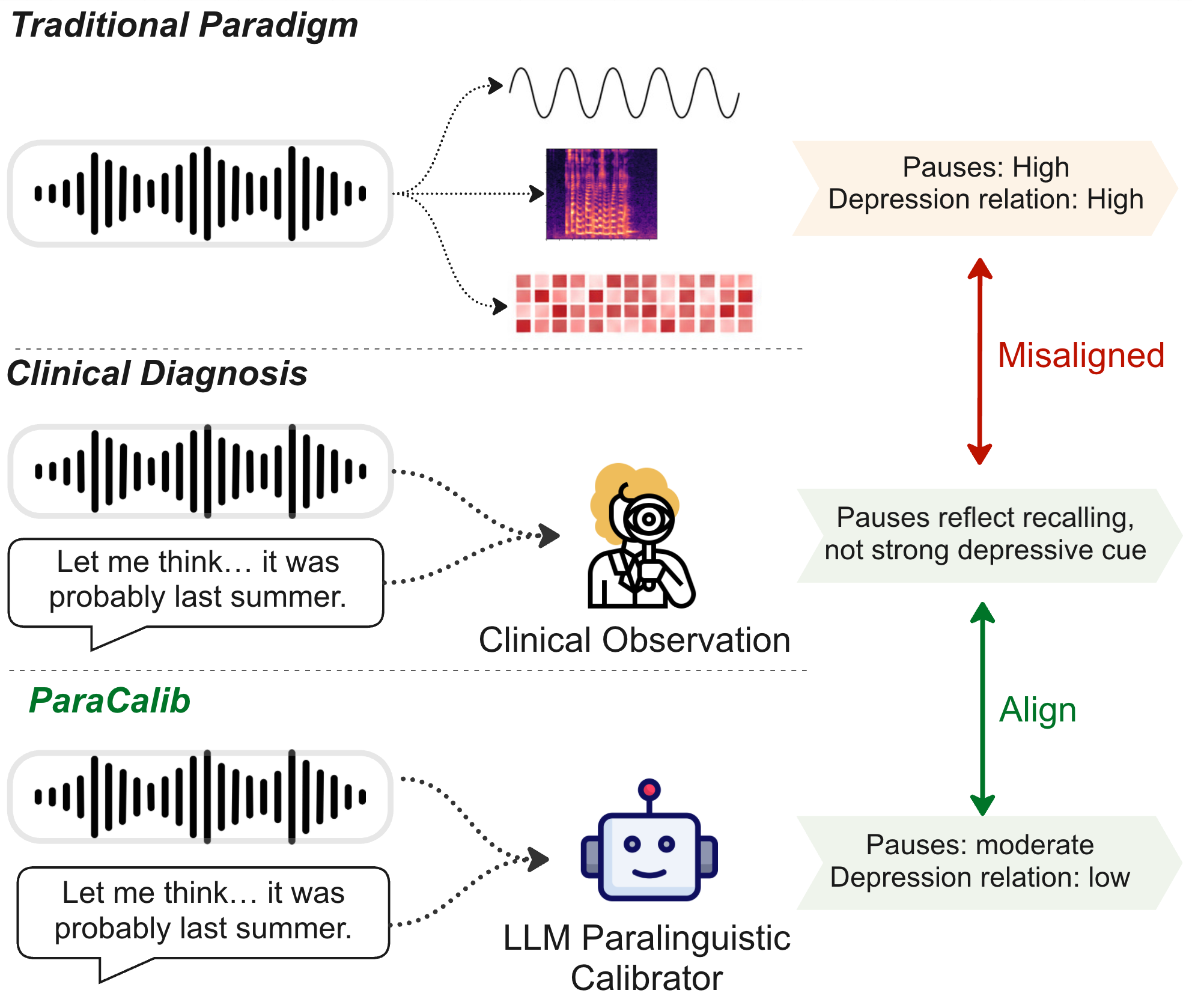}
    \caption{Illustration of the mismatch between context-independent acoustic-marker modeling and semantic calibration of paralinguistic cues. The same acoustic pattern can receive different interpretations depending on utterance-level semantic context and inferred communicative function.}
    \label{fig:intro}
\end{figure}

Depression is a prevalent mental disorder associated with substantial disability and impaired daily functioning~\citep{who2025depression}. Early and scalable screening is central to computational mental health research~\citep{galatzer2023machine}. Speech is a promising modality for this purpose because it is non-invasive, relatively easy to collect during clinical interviews or remote interactions, and psychologically informative, capturing both what a person says and how they say it~\citep{cummins2015review,li2025automated}.

Most speech-based depression detection methods follow what we refer to as an acoustic-marker paradigm: they seek vocal patterns that are statistically associated with depression and use them as participant-level indicators. Early approaches rely on low-level descriptors (LLDs), including prosodic, spectral, voice-quality, and temporal features, motivated by clinical links between depression and psychomotor slowing, longer pauses, reduced vocal energy, monotonic prosody, and altered voice quality~\citep{ellgring1996vocal,cummins2015review,yamamoto2020speech}. More recent end-to-end and self-supervised (SSL) models learn representations from raw waveforms, time-frequency inputs, or large unlabeled speech corpora~\citep{ma2016depaudionet,chen2022speechformer,baevski2020wav2vec,hsu2021hubert,Chen_2022,wu2023self,li2025hierarchical}. Despite their architectural differences, these approaches generally map acoustic form directly to depression-related evidence, without explicitly modeling how utterance-level semantic context and inferred communicative function shape the interpretation of vocal behavior.

This mismatch arises for two reasons. First, depression is clinically heterogeneous: major depressive episodes may involve either psychomotor retardation or agitation, suggesting that depression-related speech need not follow a single monotonic acoustic profile~\citep{dsm5tr2022,alpert2001reflections}. For example, depressive speech may be slower and more paused in some cases, but tense or accelerated in others. Second, the same acoustic form can serve different communicative functions. A long pause, flattened pitch contour, reduced intensity, or slowed speaking rate may reflect depressive psychomotor changes, but may also result from hesitation, emphasis, syntactic planning, turn-taking, topic shifts, or the semantic content being expressed~\citep{cole2015prosody,wagner2010prosody}. In short, depression can manifest through different vocal patterns, while the same vocal pattern can serve different communicative functions. As illustrated in Figure~\ref{fig:intro}, an acoustic pattern that appears strongly depression-related under the traditional paradigm may receive a different interpretation when considered together with the utterance and its communicative context.

Motivated by this observation, we propose ParaCalib,\footnote{Code is available at \url{https://github.com/yifeili-13/ParaCalib}.} a framework for representing utterance-level vocal behavior as semantically calibrated paralinguistic states. We define semantic calibration as interpreting vocal behavior relative to utterance-level semantic context and inferred communicative function and expressing the resulting interpretation in a structured, comparable state space. ParaCalib operationalizes this process by combining ALM-generated contextualized vocal descriptions with PSE mapping into predefined SC-Para dimensions. For each participant, the valid utterance-level SC-Para vectors form an instance bag, which is aggregated by an attention-based MIL classifier for participant-level prediction.

Experiments on DAIC-WOZ and MODMA show that ParaCalib achieves the highest mean Macro-F1 among the evaluated methods, outperforming conventional acoustic and SSL speech representations while providing inspectable intermediate paralinguistic states. Controlled analysis further shows that, under a fixed caption-level acoustic description, the PSE-derived evidence changes with semantic context, providing direct evidence that semantic calibration is operationalized at the PSE stage. Exploratory analysis identifies specific combinations of states associated with depression labels. Together, these findings support structured, context-sensitive, and compositional modeling of vocal behavior for speech-based depression detection.
\begin{figure*}[t]
    \centering
    \includegraphics[width=1\linewidth]{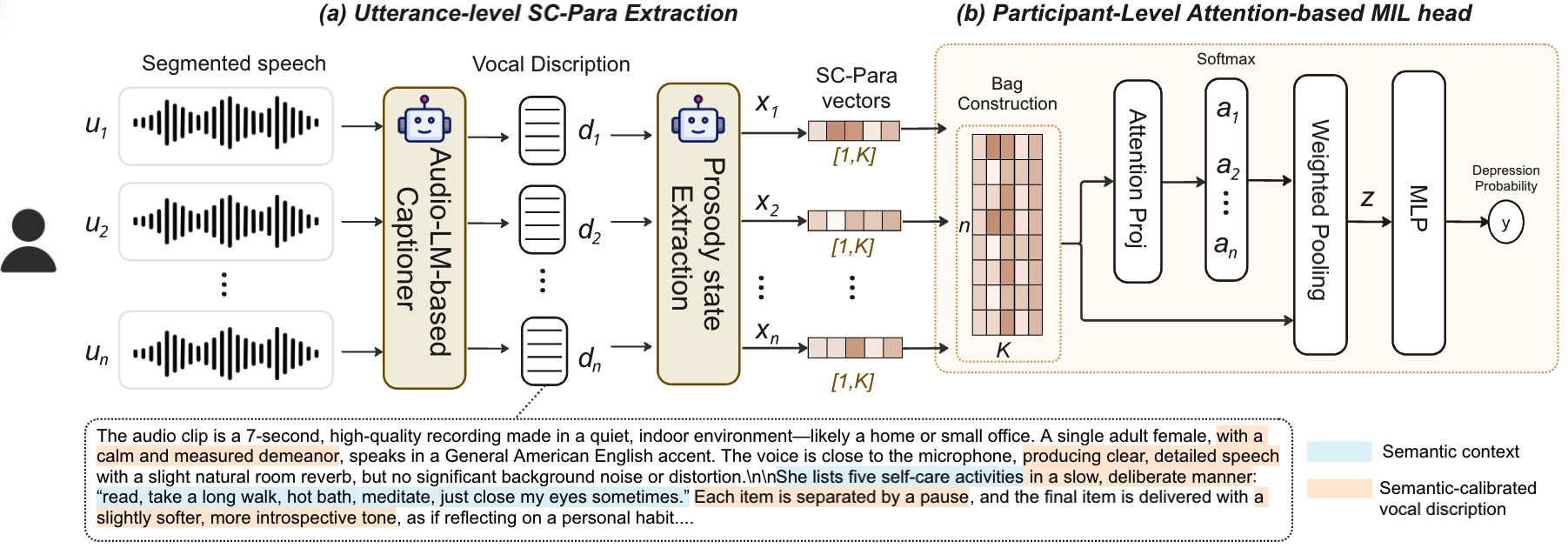}
    \caption{Overview of ParaCalib. A frozen ALM captioner converts utterance-level speech into contextualized vocal descriptions, and PSE structures those descriptions into utterance-level SC-Para vectors. The attention-based MIL classifier uses an attention projection only to estimate utterance-level instance weights, forms the participant-level representation as a weighted sum of the original seven-dimensional SC-Para vectors, and feeds this pooled representation to an MLP for depression-label prediction.}
    \label{fig:framework}
\end{figure*}     

\section{Related Work}

\subsection{Depression Detection from Speech}

The AVEC challenges have established widely used benchmarks for computational depression assessment~\citep{valstar2013avec,10.1145/3347320.3357688}. Early systems primarily relied on handcrafted prosodic, spectral, temporal, and voice-quality descriptors~\citep{li2025automated}. Their dependence on manually selected features motivated CNN-, RNN-, and Transformer-based models that learn representations directly from waveforms or time--frequency inputs~\citep{chlasta2019automated,dubagunta2019learning,saidi2020hybrid,Othmani2021,vazquez2020automatic,wang2022ecapa,Salekin2018,zhao2020hierarchical,muzammel2020audvowelconsnet}.

Pretrained speech encoders, including wav2vec~2.0~\citep{baevski2020wav2vec}, HuBERT~\citep{hsu2021hubert}, WavLM~\citep{Chen_2022}, and Whisper~\citep{radford2023robust}, further improve representation transfer in data-limited settings~\citep{Zhang2021Depa,Toto2021,chen2022speechformer,ravi2022step,wu2023self,li2025hierarchical,zhang2025speecht}. Despite their different architectures, these approaches generally use learned representations as generic features for participant-level prediction rather than explicitly modeling how vocal behavior should be interpreted in context.

Several studies incorporate linguistic information or examine its potential confounding effects. Syntactic information has been used to guide acoustic feature extraction and multimodal fusion~\citep{morales2018linguistically}. DepFlow addresses semantic shortcut reliance by synthesizing examples that decouple depression-conditioned vocal expression from linguistic sentiment~\citep{li2026depflow}. These studies demonstrate that linguistic content can provide relevant contextual information while also introducing spurious label associations. However, they do not explicitly represent how the meaning of a vocal pattern changes with utterance-level semantic context and inferred communicative function.

\subsection{Audio-Language Models for Affective Speech Understanding}

Audio-Language Models (ALMs), such as SALMONN, Qwen-Audio, Qwen2-Audio, and GAMA, connect audio encoders with pretrained language models to support general audio understanding~\citep{tang2024salmonn,chu2023qwenaudioadvancinguniversalaudio,chu2024qwen2audiotechnicalreport,ghosh2024gama}. In affective speech, these models have been applied to closed-set emotion recognition, free-form vocal captioning, and prediction refinement grounded in transcripts and acoustic cues~\citep{zhang-etal-2025-beyond-classification,xu2024secap}. Related work has also explored acoustic-grounded reasoning~\citep{tian2025stepaudior1technicalreport}, ordinal modeling of affective dimensions~\citep{chandra2025emotionrankclap}, and multimodal LLMs for depression assessment~\citep{zhang2024llms,zhang2025speecht,zhao2025hears,CHEN2026113367}.

These approaches show that ALMs can generate semantically rich descriptions of vocal behavior and support flexible affective reasoning. However, free-form descriptions are difficult to compare consistently across utterances and participants, while closed-set affective classifiers often compress vocal behavior into coarse categories. Existing systems primarily optimize prediction or explanation generation rather than constructing a structured representation that preserves context-dependent paralinguistic distinctions.

ParaCalib addresses this gap by mapping contextualized vocal descriptions into fixed Semantically Calibrated Paralinguistic (SC-Para) dimensions. This design combines the contextual sensitivity of ALM-generated descriptions with the comparability of a structured state space. Unlike generic acoustic representations, SC-Para explicitly conditions the interpretation of vocal behavior on utterance-level semantic context and inferred communicative function. Unlike free-form captions or explanations, it supports participant-level aggregation and representation-level analysis.

\section{Methodology}
\label{sec:methodology}

\subsection{Overview}

As illustrated in Figure~\ref{fig:framework}, ParaCalib consists of three stages: ALM-based vocal captioning, PSE-based SC-Para extraction, and attention-based MIL aggregation. Given an utterance-level audio segment, the frozen ALM generates a natural-language description that captures both the spoken content and the accompanying vocal behavior. The frozen PSE then converts this description into a fixed-dimensional SC-Para vector. Finally, the attention-based MIL classifier aggregates utterance-level SC-Para vectors into a participant-level representation for depression-label prediction. The ALM and PSE remain frozen throughout training, and only the downstream MIL classifier is optimized.

\subsection{Utterance-level ALM Captioning}

Let $u_i$ denote the $i$-th utterance-level audio segment. We use a frozen Qwen3-Omni-30B-A3B Captioner~\citep{xu2025qwen3omnitechnicalreport}\footnote{\url{https://huggingface.co/Qwen/Qwen3-Omni-30B-A3B-Captioner}} to generate a natural-language vocal description:
\begin{equation}
    d_i = \mathrm{ALM}(u_i).
\end{equation}

The ALM processes the audio input as a unified audio-language signal. It can infer both what is being said and how it is being said, allowing it to describe vocal behavior in relation to the utterance content. The generated description may include observable properties such as pitch variability, speaking rate, rhythm, pause pattern, vocal energy, and voice quality, together with their interpretation in the context of the utterance.

The captioner receives only the utterance-level audio; no external text prompt is supplied because the captioning instruction is built into the specialized checkpoint. The ALM is not fine-tuned on either dataset and does not receive participant-level depression labels. The checkpoint and decoding settings are provided in Appendix~\ref{app:model_settings}.

\subsection{SC-Para Extraction}

Although ALM-generated descriptions provide rich information about vocal behavior, their free-form language is difficult to compare consistently across utterances and participants. We therefore use a frozen DeepSeek-R1-Distill-Qwen-32B model~\citep{Guo_2025}\footnote{\url{https://huggingface.co/deepseek-ai/DeepSeek-R1-Distill-Qwen-32B}} as the PSE to convert each description $d_i$ into a fixed SC-Para schema:
\begin{equation}
    x_i = \mathrm{PSE}(d_i),
\end{equation}
where $x_i \in \mathbb{R}^{K}$ and $K=7$.

In the main evaluation, the frozen PSE receives only the ALM-generated utterance description; it is not provided with participant identifiers, PHQ scores, or dataset labels.

The SC-Para schema contains five constituent paralinguistic dimensions and two holistic dimensions. The five constituent dimensions describe pitch variability, speaking rate, rhythm regularity, pause pattern, and vocal energy. These dimensions reflect paralinguistic phenomena commonly examined in clinical and affective speech research~\citep{cummins2015review,koops2023speech}. Each constituent dimension is assigned one of three discrete states and encoded numerically as $0$, $1$, or $2$. The two holistic dimensions are a depression-related paralinguistic evidence score and a confidence score. The evidence score is generated on a $0$--$100$ scale and normalized to $[0,1]$, while the confidence score is represented directly on $[0,1]$. These scores are PSE-generated intermediate variables rather than utterance-level gold depression labels, and provide additional information derived from the PSE's pretrained linguistic and affective knowledge. Their contribution is evaluated through component-level ablations and prompt-sensitivity analyses in Appendix~\ref{app:ablations}.

For each utterance $u_i$, the PSE produces
\begin{equation}
    x_i =
    [x_i^{(1)},x_i^{(2)},\ldots,x_i^{(K)}].
\end{equation}

If the PSE assigns \texttt{unknown} to any categorical dimension because the description provides insufficient vocal evidence, the utterance is excluded before participant-level aggregation. The two holistic scores are retained for all utterances with valid categorical states. Discard rates are reported in Table~\ref{tab:discard} of Appendix~\ref{app:robustness}. The PSE prompt template and an abbreviated output schema are provided in Appendix~\ref{app:prompts}.

\subsection{Attention-based MIL Classification}

For participant $b$, let
\begin{equation}
    S_b=\{x_{b,1},x_{b,2},\ldots,x_{b,n_b}\}
\end{equation}
denote the set of valid utterance-level SC-Para vectors, where $n_b$ may vary across participants. Each participant is treated as a bag of utterance-level instances.

The attention-based MIL classifier assigns an importance score to each instance using an additive attention function:
\begin{equation}
    e_{b,i}
    =
    v_{\mathrm{att}}^{\top}
    \tanh
    \left(
        W_{\mathrm{att}}x_{b,i}
        +b_{\mathrm{att}}
    \right),
\end{equation}
where $W_{\mathrm{att}}$, $b_{\mathrm{att}}$, and $v_{\mathrm{att}}$ are learnable parameters. The attention weights are normalized across the utterances belonging to the same participant:
\begin{equation}
    \alpha_{b,i}
    =
    \frac{\exp(e_{b,i})}
    {\sum_{j=1}^{n_b}\exp(e_{b,j})}.
\end{equation}

The utterance-level vectors are pooled into a participant-level representation:
\begin{equation}
    z_b
    =
    \sum_{i=1}^{n_b}
    \alpha_{b,i}x_{b,i}.
\end{equation}

The attention module estimates the relative contribution of each utterance, while the readout preserves the original seven-dimensional SC-Para representation. The pooled representation $z_b$ is then passed to a two-hidden-layer multilayer perceptron (MLP), which produces a scalar logit $s_b$ for binary depression-label classification:
\begin{equation}
    \hat{y}_b = \sigma(s_b),
\end{equation}
where $\sigma(\cdot)$ denotes the sigmoid function. The architecture and training hyperparameters are reported in Appendix~\ref{app:setup}, Table~\ref{tab:hyperparams}.

\subsection{Optimization Objective}

The ALM and PSE remain frozen, and only the attention module and
downstream MLP are optimized. We use weighted binary cross-entropy
with positive-class weight
$w_{+}=N_{\mathrm{neg}}/N_{\mathrm{pos}}$ and
negative-class weight $w_{-}=1$. The remaining optimization settings
are provided in Appendix~\ref{app:setup}.

\begin{table*}[t]
\centering
\small
\setlength{\tabcolsep}{2pt}
\begin{tabular}{lllcccc}
\toprule
Dataset & Method & Input &  Macro-F1 ($\uparrow$) & Sensitivity ($\uparrow$) & Specificity ($\uparrow$) & AUC ($\uparrow$) \\
\midrule
\multirow{6}{*}{DAIC-WOZ} 
& DepAudioNet~\citep{ma2016depaudionet}\textsuperscript{*} & MFbanks 
& 50.7 (1.3) & 67.1 (8.1) & 44.8 (5.1)& 54.1 (2.6) \\
& NUSD~\citep{wang2023non}\textsuperscript{*} & Raw speech 
& 51.4 (4.6) & 30.0 (13.7) & 74.5 (10.2) & 57.5 (3.4) \\
& Wu et al.-style~\citep{wu2023self}\textsuperscript{*} & WavLM 
& 53.5 (5.3) & 32.8 (24.0) & 78.2 (18.0) & 54.7 (7.5) \\
& SpeechFormer~\citep{chen2022speechformer}\textsuperscript{*} & Wav2Vec 1.0 
& 	
58.0 (3.5) & 40.0 (3.9) & 75.8 (4.3) & 54.1 (2.4) \\
& HAREN-CTC~\citep{li2025hierarchical}\textsuperscript{*} & WavLM 
& 57.9 (2.1) & 41.4 (12.0) & \textbf{83.6} (7.0) & 55.6 (1.0) \\
& ParaCalib & SC-Para 
& \textbf{71.9} (4.9) & \textbf{77.1} (15.5) & 73.3 (6.6) & \textbf{83.3} (2.6) \\
\midrule
\multirow{4}{*}{MODMA}
& Wu et al.-style~\citep{wu2023self}\textsuperscript{*} & WavLM 
& 59.7 (4.9) & 50.0 (14.0) & 70.0 (6.2)& 64.8 (6.6)\\
& DepAudioNet~\citep{ma2016depaudionet}\textsuperscript{*} & MFbanks 
& 69.5 (6.3) & 52.7 (10.0) & 86.0 (4.9) & 74.7 (6.0) \\
& HAREN-CTC~\citep{li2025hierarchical}\textsuperscript{*} & WavLM
& 73.1 (5.9) & 61.4 (8.3) & 86.0 (9.2) & 76.4 (7.6) \\
& ParaCalib & SC-Para 
& \textbf{90.5} (11.2) & \textbf{91.0} (12.5) & \textbf{90.0} (14.9) & \textbf{96.4} (6.9) \\
\bottomrule
\end{tabular}
\caption{
Main participant-level depression-label classification results on DAIC-WOZ and MODMA.
\textsuperscript{*} indicates reproduced baselines. Open-source baselines use the released implementations and reported configurations. Wu et al.-style denotes pretrained WavLM without the sub-dialogue augmentation used in the original study. All reproduced methods follow our common preprocessing, model-selection, and held-out test protocol. Results are reported as mean (SD) in percentage points.
}
\label{tab:main_results}
\end{table*}

\begin{table*}[t]
\centering
\small
\setlength{\tabcolsep}{2pt}
\begin{tabular}{lllcccc}
\toprule
Dataset & Representation type & Representation 
& Macro-F1 ($\uparrow$) & Sensitivity ($\uparrow$) 
& Specificity ($\uparrow$) & AUC ($\uparrow$) \\
\midrule
\multirow{11}{*}{DAIC-WOZ}
& \multirow{2}{*}{LLD-based}
& LLDs 
& 43.9 (6.4) & 35.7 (29.0) & \cellcolor{blockbest}62.4 (34.9) & 48.1 (4.4) \\
& 
& Discretized LLDs
& \cellcolor{blockbest}49.0 (6.2) & \cellcolor{blockbest}40.0 (21.2) & 61.2 (14.4) & \cellcolor{blockbest}50.4 (6.9) \\
\cmidrule(lr){2-7}
& \multirow{3}{*}{SSL-based}
& Wav2Vec 2.0
& 41.8 (10.5) & \cellcolor{blockbest}40.0 (25.6) & 53.3 (33.2) & 42.3 (4.6) \\
&
& HuBERT
& \cellcolor{blockbest}46.3 (5.9) & 37.1 (12.8) & 57.0 (6.6)& 40.1 (3.5)\\
&
& WavLM
& 41.1 (7.3) & 28.6 (34.6) & \cellcolor{blockbest}66.7 (36.9)  &\cellcolor{blockbest}46.6 (5.7) \\
\cmidrule(lr){2-7}
& \multirow{3}{*}{Transcript-based}
& Transcript 
& \cellcolor{blockbest}63.1 (13.3) & \cellcolor{blockbest}52.9 (30.1) & 79.4 (12.6) & \cellcolor{blockbest}73.5 (3.2) \\
& 
& LLD-PSE w/ transcript
& 54.5 (8.2) & 28.6 (18.9) & \cellcolor{blockbest}\textbf{84.2} (15.4) & 64.2 (6.6) \\
&
& Transcript + WavLM
& 46.1 (11.1) & 20.0 (22.2) &81.2 (29.2) & 49.7 (12.4) \\
\cmidrule(lr){2-7}
& \multirow{3}{*}{Caption-based}
& Transcript-masked Caption-PSE
& 64.6 (3.8) & 51.4 (9.3) & \cellcolor{blockbest}78.2 (7.8) & 77.9 (2.0) \\
& 
& Caption Embedding
& 66.6 (14.2) & 68.6 (39.3) & 75.8 (14.1) & 77.1 (12.0) \\
& 
& SC-Para
& \cellcolor{blockbest}\textbf{71.9} (4.9) & \cellcolor{blockbest}\textbf{77.1} (15.5) & 73.3 (6.6) & \cellcolor{blockbest}\textbf{83.3} (2.6) \\
\midrule
\multirow{8}{*}{MODMA}
& \multirow{2}{*}{LLD-based}
& LLDs 
& 53.4 (11.5) & 35.4 (14.1) & 79.1 (12.3) & 57.8 (13.4) \\
& 
& Discretized LLDs
& \cellcolor{blockbest}67.8 (19.1) & \cellcolor{blockbest}56.4 (25.6) & \cellcolor{blockbest}82.0 (11.9) & \cellcolor{blockbest}78.7 (19.4) \\
\cmidrule(lr){2-7}
& \multirow{3}{*}{SSL-based}
& Wav2Vec 2.0
& 55.6 (7.7) & 62.6 (24.0) & 59.2 (16.8) & 63.7 (7.4) \\
&
& HuBERT
& 61.6 (24.7) & 62.8 (19.1) & 75.1 (18.3) & 71.8 (22.3)\\
&
& WavLM
& \cellcolor{blockbest}68.8 (20.5) & \cellcolor{blockbest}63.2 (25.8) & \cellcolor{blockbest}79.1 (15.9) & \cellcolor{blockbest}76.5 (19.6) \\
\cmidrule(lr){2-7}
& \multirow{3}{*}{Caption-based}
& Transcript-masked Caption-PSE
& 60.6 (19.8) & 58.2 (14.9) & 71.9 (17.0) & 69.2 (13.3) \\
& 
& Caption Embedding
& 85.7 (10.0) & 86.4 (9.0) & 85.3 (11.0) & 94.3 (7.0) \\
& 
& SC-Para
& \cellcolor{blockbest}\textbf{90.5} (11.2) & \cellcolor{blockbest}\textbf{91.0} (12.5) & \cellcolor{blockbest}\textbf{90.0} (14.9) & \cellcolor{blockbest}\textbf{96.4} (6.9) \\
\bottomrule
\end{tabular}
\caption{
Representation analysis of ParaCalib across alternative representation strategies. Results are reported as mean (SD) in percentage points. Shaded cells indicate the best performance within each representation type, while bold values denote the overall best result per dataset and metric. Transcript-based baselines are omitted for MODMA due to the lack of available transcripts. Because the representations differ in upstream pretraining, dimensionality, and extraction complexity, this table compares complete representation strategies under a common downstream evaluation protocol rather than parameter-matched models. Additional extraction configurations and dimension specifications are given in Appendix~\ref{app:setup}.
}
\label{tab:semantic_pse_analysis}
\end{table*}

\section{Data and Evaluation}
\label{sec:data_evaluation}

\subsection{Datasets}

We evaluate ParaCalib on two speech-based depression datasets: DAIC-WOZ~\citep{gratch-etal-2014-distress} and MODMA~\citep{Cai_2022}.

DAIC-WOZ contains 189 English-speaking participants and provides an official participant-disjoint training, development, and test split. We define depression-label status using a PHQ-8 threshold of 10~\citep{kroenke2009phq}. Following prior preprocessing~\citep{bailey2021gender}, we correct the label assignment of participant 409. The resulting cohort contains 57 depression-label and 132 non-depression-label participants.

MODMA contains 52 Mandarin-speaking participants, with 18 interview utterances per participant. Although the original participant groups were established through psychiatric assessment, including the MINI under DSM-IV criteria, our experiments use PHQ-9-defined depression labels with a threshold of 10~\citep{kroenke2001phq}. Following prior preprocessing~\citep{li2025hierarchical}, this criterion reassigns participant 2010037 to the non-depression-label group, resulting in 22 depression-label and 30 non-depression-label participants. Thus, the MODMA results concern PHQ-9-based labels rather than the original diagnostic grouping. Label definitions and preprocessing details are provided in Appendix~\ref{app:setup}.

\subsection{Evaluation Protocol}

For DAIC-WOZ, we use the official training, development, and test partitions. For MODMA, we perform participant-level five-fold cross-validation and repeat each fold with the same five random seeds. Within each fold, all preprocessing, model selection, and classifier training are conducted without access to the held-out participants.

We report participant-level Macro-F1 as the primary metric, together with sensitivity, specificity, and area under the receiver operating characteristic curve (AUC). Unless otherwise stated, thresholded predictions use a fixed decision threshold of 0.5. All results are reported as mean and standard deviation (SD) in percentage points. Specifically, DAIC-WOZ results are aggregated over five matched random seeds. For MODMA, metrics are first averaged over the five seeds within each fold, and the reported mean and SD are then computed across the five participant-level fold means. Reproduced acoustic, end-to-end, and self-supervised baselines, including SpeechFormer, use the same participant-level splits, preprocessing procedure, model-selection protocol, and held-out evaluation protocol~\citep{ma2016depaudionet,wang2023non,wu2023self,chen2022speechformer,li2025hierarchical}. Implementation and model settings are provided in Appendix~\ref{app:setup}.

\subsection{Statistical and Diagnostic Analyses}

Representation comparisons evaluate complete representation strategies under the same attention-based MIL classifier and training protocol. For each comparison, we conduct a two-sided paired $t$-test across the five matched evaluation units defined above and report the 95\% confidence interval for the mean paired difference. These comparisons characterize variation across matched runs or fold-level evaluation units rather than uncertainty obtained by resampling participants. When multiple comparisons are performed within the same analysis family, we apply Holm correction to the corresponding $p$-values. Additional details of the statistical tests are provided in Appendix~\ref{app:setup}.

We perform several analyses to examine the properties of SC-Para. First, the semantic-context control holds the caption-level acoustic description fixed while varying the accompanying context. We compare three conditions: acoustic-only context, non-depression-related context, and depression-related context. The analysis uses 100 utterances from 66 participants. The construction of the context conditions and the corresponding inferential procedure are described in Appendix~\ref{app:analysis_details}.

Second, prompt sensitivity is evaluated on the same 100 captions using two meaning-preserving prompt paraphrases, while keeping the SC-Para schema, numerical mappings, and output format fixed. Third, we conduct an exploratory utterance-level contingency-table analysis of associations between categorical SC-Para states and depression labels. Because multiple utterances share the same participant label, the adjusted significance values are used only to screen descriptive patterns and are not interpreted as participant-level confirmatory inference. The construction of the contingency tables and the screening criteria are described in Appendix~\ref{app:analysis_details}.

We additionally evaluate speaker-identity leakage using a participant-disjoint speaker-verification probe and assess attribute-wise removal through retrained MIL classifiers. Finally, additive-noise robustness is evaluated by mixing background-noise recordings from the MUSAN noise partition~\citep{snyder2015musan} into DAIC-WOZ test utterances at 20, 10, and 5 dB signal-to-noise ratios. These perturbations are applied only during robustness evaluation and do not affect model training or model selection. Additional results are reported in Appendices~\ref{app:ablations} and~\ref{app:robustness}.

\section{Results and Analysis}

\subsection{Depression Detection Performance}

Table~\ref{tab:main_results} reports participant-level classification results on DAIC-WOZ and MODMA. ParaCalib achieves the highest mean Macro-F1 among the evaluated methods on both datasets, reaching 71.9\% on DAIC-WOZ and 90.5\% on MODMA. It also obtains the highest sensitivity and AUC on both datasets. Relative to HAREN-CTC, ParaCalib improves mean Macro-F1 by
14.0 percentage points on DAIC-WOZ and 17.5 percentage points on MODMA, based on the unrounded matched-unit means. The paired evaluation-unit comparisons yield Holm-adjusted $p$-values of .0232 and .0734, respectively (Table~\ref{tab:sig}). Split-wise results and exploratory severity-stratified analyses are reported in Appendix~\ref{app:setup} and Appendix~\ref{app:robustness}.

\subsection{Representation Comparison}

\label{sec:representation_analysis}

Table~\ref{tab:semantic_pse_analysis} compares complete representation strategies under the same attention-based MIL evaluation protocol. Discretizing eGeMAPS descriptors improves Macro-F1 from 43.9\% to 49.0\% on DAIC-WOZ and from 53.4\% to 67.8\% on MODMA. The strongest generic SSL representations reach 46.3\% and 68.8\%, respectively, whereas SC-Para reaches 71.9\% and 90.5\%.
 
On DAIC-WOZ, transcript-based representations achieve Macro-F1 values between 46.1\% and 63.1\%. Transcript-masked Caption-PSE reaches 64.6\% on DAIC-WOZ and 60.6\% on MODMA, indicating that vocal-delivery information remains predictive after explicit lexical content is removed. The larger reduction under content masking on MODMA indicates a stronger contribution from semantic context on that dataset. Accordingly, ParaCalib integrates contextual semantic and vocal-delivery information rather than isolating a purely acoustic marker. The content-masking prompt is provided in Appendix~\ref{app:prompts}.

Direct Caption Embedding reaches 66.6\% Macro-F1 on DAIC-WOZ and
85.7\% on MODMA. SC-Para further improves the mean Macro-F1 to 71.9\%
and 90.5\%, corresponding to gains of 5.3 and 4.8 percentage points,
respectively; the Holm-adjusted paired comparisons yield
$p_{\mathrm{adj}}=.7269$ on both datasets. Across both datasets,
SC-Para therefore achieves higher mean performance while converting
free-form captions into a compact, fixed-dimensional, and inspectable
representation. This pattern suggests that structuring contextualized
vocal descriptions into semantically calibrated states preserves and
may sharpen task-relevant information while enabling dimension-level
analysis.

\subsection{Semantic-Context Control}

We next examine whether semantic context affects PSE interpretation
independently of the vocal-behavior description. The controlled analysis
holds this description fixed and compares acoustic-only,
non-depression-related, and depression-related context conditions.
Depression-related context increases the PSE evidence score by 8.6
points relative to non-depression-related context
(participant-clustered 95\% CI $[6.7,10.6]$). The corresponding
increase relative to acoustic-only input is 8.7 points, whereas
non-depression-related context differs from acoustic-only input by only
0.1 points ($p=.92$). These results provide direct evidence that
semantic calibration is operationalized at the PSE stage: the same
caption-level acoustic description yields different depression-related
paralinguistic evidence under different semantic contexts. Additional
test statistics are provided in Appendix~\ref{app:analysis_details}.

 \subsection{Component and State-Level Analysis}

Attentive MIL achieves the highest Macro-F1 among the evaluated participant-level classifiers on both datasets; the complete classifier comparison is provided in Appendix~\ref{app:setup}.

Figure~\ref{fig:feature_importance} summarizes the metric-wise performance changes after removing each SC-Para dimension and retraining the attentive MIL classifier on DAIC-WOZ. The largest Macro-F1 decreases occur after removing the PSE-derived evidence score (5.0 percentage points), speaking rate (4.7 points), and vocal energy (4.5 points). Their effects differ across metrics: vocal energy and the evidence score contribute primarily to sensitivity, whereas speaking rate contributes more strongly to specificity.

Beyond individual-dimension removal, the five categorical dimensions alone achieve 64.0\% Macro-F1, while the two holistic scores alone achieve 64.7\% and the evidence score alone achieves 63.2\%, compared with 71.9\% for the complete representation. Replacing the depression-specific holistic scores with a generic affect/arousal prompt reduces Macro-F1 from 64.7\% to 61.3\% and AUC from 69.6\% to 61.9\%. These results suggest that the categorical and holistic dimensions provide complementary predictive information.

Pairwise Spearman correlations among the five categorical dimensions range from $-0.400$ to $0.703$ (Table~\ref{tab:corr}), indicating that the dimensions are related but not redundant.

\begin{figure}[t]
    \centering
    \includegraphics[width=\linewidth]{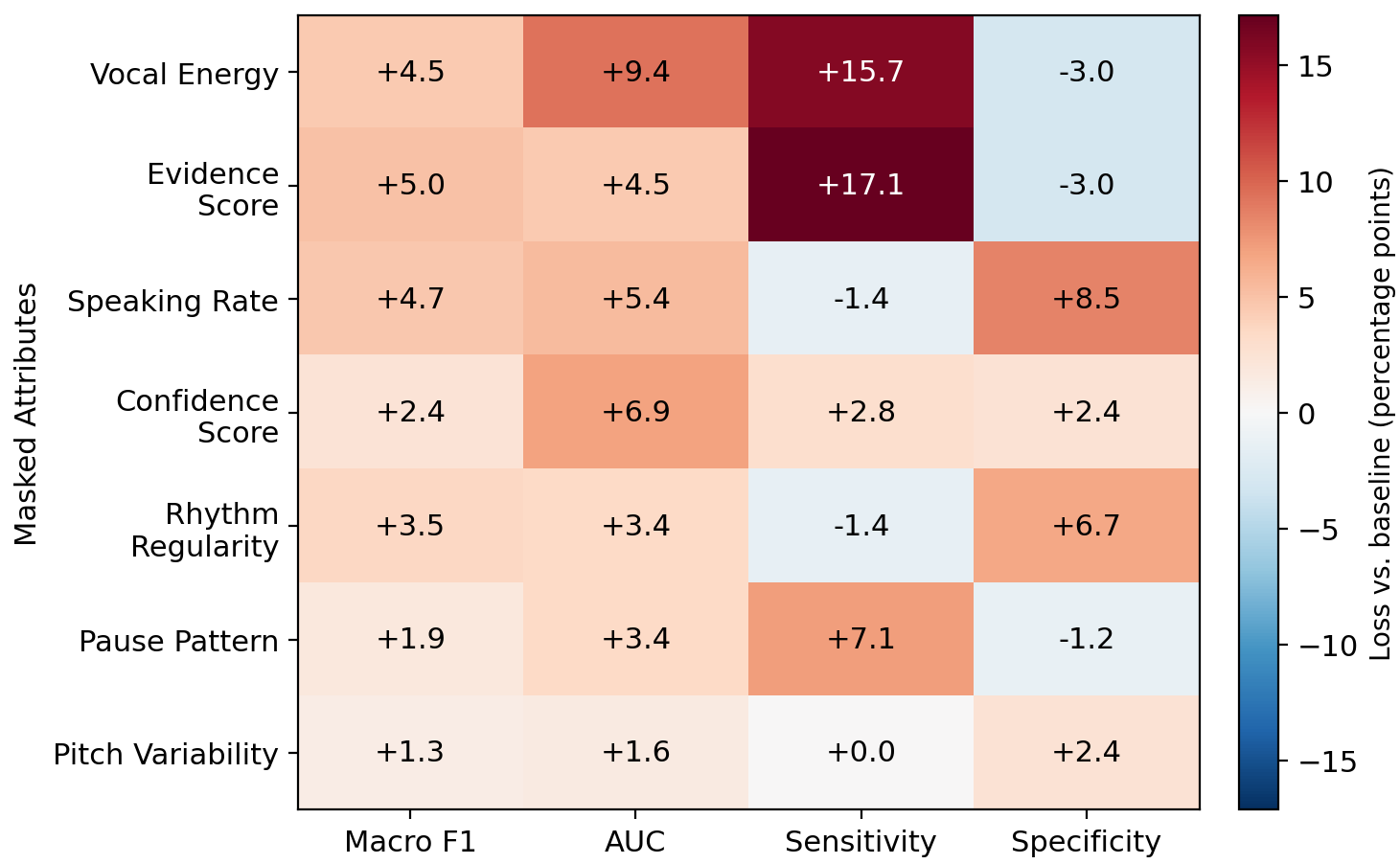}
    \caption{Metric-wise performance changes on DAIC-WOZ after removing one SC-Para dimension and retraining the attentive MIL classifier. Values are calculated as the five-seed mean performance of the full model minus that of the corresponding ablated model and are reported in percentage points. Positive values indicate performance degradation after removal, whereas negative values indicate improvement.}
    \label{fig:feature_importance}
\end{figure}

\paragraph{State-Level and Pairwise Associations.}

In this exploratory utterance-level analysis, marginal associations are selective rather than uniformly monotonic across states. Low energy, flat pitch, moderate pauses, and specific speaking-rate states show positive associations with depression labels relative to their corresponding reference states. Seven specific combinations of states met the prespecified frequency and multiplicity-adjusted display criteria; several involve low energy together with slow speech, steady rhythm, few pauses, or flat pitch. Because utterances are clustered within participants, these patterns are descriptive associations rather than participant-level confirmatory evidence. The displayed associations are shown in Figure~\ref{fig:interpretable}, and the screening criteria are given in Appendix~\ref{app:analysis_details}.

\begin{figure*}[t]
    \centering
    \includegraphics[width=0.9\linewidth]{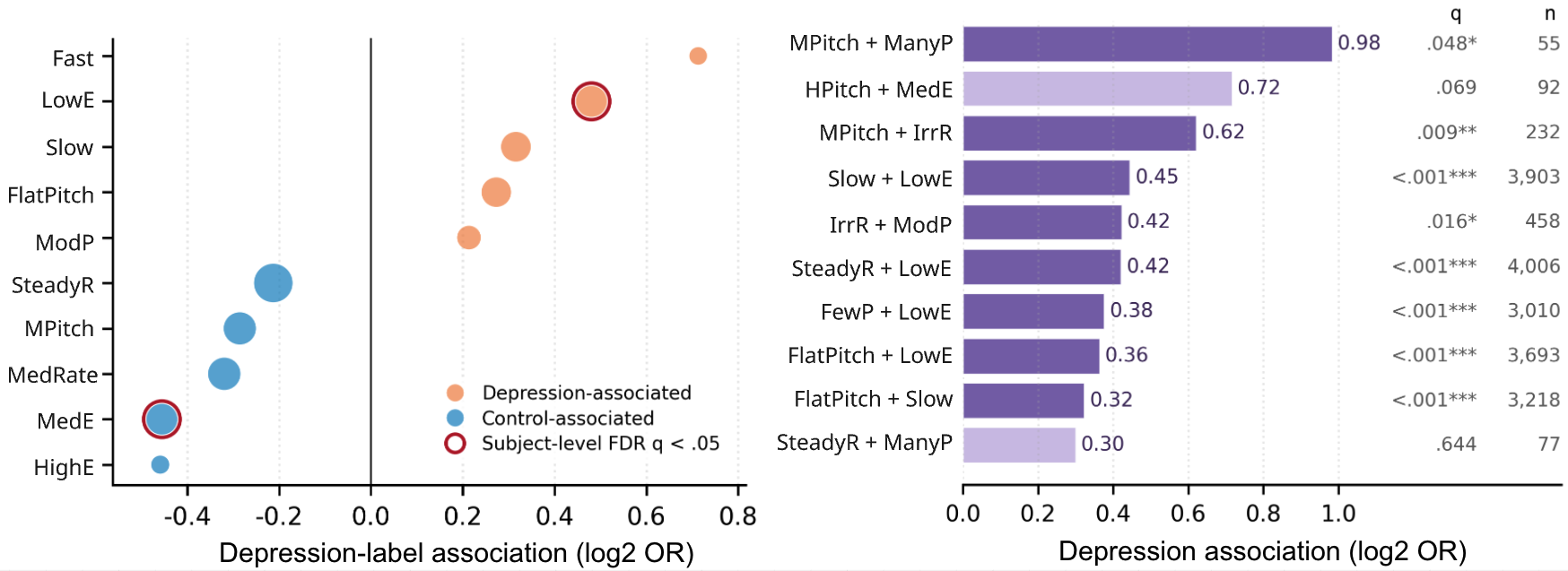}
    \caption{Exploratory utterance-level associations between categorical
SC-Para states and participant-level depression labels. Left: marginal
log$_2$ odds ratios. Right: the ten highest-ranked state pairs. We
interpret only the seven pairs satisfying $n\geq100$ and $q<.05$; the
remaining three are shown descriptively. The $q$ values do not account
for within-participant clustering.}
    \label{fig:interpretable}
\end{figure*}

\subsection{Robustness and Error Analysis}

We conduct additional analyses of speaker identity, extraction stability, acoustic robustness, and prediction errors. Additional protocols and results are reported in Appendices~\ref{app:robustness} and~\ref{app:ablations}.

A participant-disjoint speaker-verification probe obtains an AUC of 52.9\% for the full seven-dimensional SC-Para representation, compared with 78.7\% for WavLM embeddings. This result suggests that SC-Para contains substantially less speaker-identity information than the generic SSL representation. In an exploratory DAIC-WOZ test-set analysis, depression-label participants showed lower attention-weighted SC-Para dispersion than non-depression-label participants (mean difference $=-0.0327$; Appendix~\ref{app:dispersion}).

Repeated-decoding analysis yields 94--99\% exact agreement across SC-Para dimensions. Prompt paraphrase analysis also shows high consistency, with correlations of $r\geq .841$. In contrast, additive acoustic noise degrades performance: mixing background noise at 5 dB SNR reduces Macro-F1 from 71.9\% to 62.6\%.

In a small internal human validation, both raters judged 18 of 20 captions and 17 of 20 constituent-state sets as consistent with perceived vocal evidence; details are provided in Appendix~\ref{app:human_validation} and Table~\ref{tab:human_val}.

Finally, consistent false positives tend to occur near the PHQ-8 cutoff, with a mean PHQ-8 score of 6.4 compared with 2.6 for reliably classified controls. Consistent false negatives contain fewer valid utterances, averaging 66 utterances compared with 119 for reliably classified true positives. These descriptive error patterns are summarized in Appendix~\ref{app:robustness}.

\section{Discussion}
\label{sec:discussion}
The results support semantic calibration as a useful modeling principle for depression-related paralinguistic evidence. Rather than treating vocal patterns as context-independent markers, ParaCalib interprets them jointly with utterance-level semantic information and maps the resulting interpretation into a structured state space. Under a fixed caption-level acoustic description, changing the accompanying semantic context shifts the PSE-derived evidence score, demonstrating context sensitivity at the PSE stage. 

The representation comparisons further clarify the role of SC-Para. SC-Para substantially outperforms the evaluated handcrafted and SSL-based representations and achieves higher mean Macro-F1 than direct Caption Embedding on both datasets. These consistent numerical improvements indicate that the structured state representation preserves and may refine the task-relevant information contained in free-form captions. Importantly, SC-Para also provides a compact and fixed-dimensional representation whose intermediate paralinguistic dimensions can be compared across utterances and participants. The component analyses further suggest that categorical vocal states and PSE-derived holistic scores provide complementary information.

This structured representation enables predictions to be inspected at the utterance level before participant-level aggregation. State-level associations, attribute-wise removal, and attention-weighted dispersion offer complementary descriptions of the patterns used by the classifier. However, these analyses should not be interpreted as causal feature importance or as direct evidence of clinically observed symptoms. They characterize the behavior of the extracted representation and the downstream classifier under the evaluated datasets and protocols.

\section{Conclusion}
ParaCalib operationalizes semantic calibration by converting contextualized vocal descriptions into a compact and inspectable SC-Para representation for participant-level depression-label classification. It achieves the highest mean Macro-F1 among all evaluated representation strategies on both DAIC-WOZ and MODMA, yielding higher mean performance than conventional acoustic, SSL-based, transcript-based, and direct caption-embedding alternatives. Controlled analysis further shows that the PSE-derived evidence is sensitive to semantic context under a fixed caption-level acoustic description. Together, these findings support structured, context-sensitive, and compositional modeling of vocal behavior.

\section{Limitations}
\label{sec:limitations}

We note several considerations for interpreting and extending ParaCalib. First, as with other speech-based approaches, performance depends on recording conditions and the amount of available speech. Performance decreases under severe additive noise, and participants with fewer valid utterances provide less evidence for participant-level aggregation. These observations motivate future work on noise-robust extraction and uncertainty-aware prediction.

Second, our experimental targets are binary labels derived from PHQ-8 and PHQ-9 cutoffs. Accordingly, the reported results should be interpreted as depression-label prediction rather than clinical diagnosis. Further validation on prospectively collected data and independently adjudicated clinical outcomes would be needed before considering clinical deployment.

Third, ParaCalib uses frozen generative models for captioning and state extraction. Although the extraction is stable under repeated decoding and prompt paraphrasing, future work should examine robustness across alternative ALMs, PSEs, recording conditions, and prompting choices. The current \texttt{unknown}-filtering procedure also warrants continued evaluation; its discard rates are reported in Table~\ref{tab:discard}.

Fourth, the state-association analysis is exploratory because multiple utterances from the same participant share one label. The displayed $q$ values do not account for this within-participant clustering and therefore cannot be interpreted as participant-level confirmatory significance. A participant-clustered or participant-level reanalysis is needed to confirm these patterns.

Finally, DAIC-WOZ and MODMA are evaluated independently. The consistent results across these datasets do not constitute a cross-dataset or cross-lingual transfer evaluation. Moreover, the paired comparisons in Table~\ref{tab:sig} are based on five matched runs or fold-level means and therefore characterize evaluation-unit variation rather than participant-resampling uncertainty. Larger and more diverse cohorts would enable stronger analyses of demographic robustness, generalizability, and real-world screening utility.

\section*{Ethics Statement}

This study follows the ACL Ethics Policy and uses existing speech datasets for computational mental health research.

\paragraph{Data Access and Original Consent}
We obtained DAIC-WOZ~\citep{gratch-etal-2014-distress} and MODMA~\citep{Cai_2022} through their official distribution channels and used them in accordance with the applicable research-access conditions. The original dataset documentation describes the corresponding ethical-approval and informed-consent procedures. Our study involved no new participant recruitment, participant contact, or prospective data collection.

\paragraph{Privacy and Data Protection}
Although direct identifiers had been removed by the dataset providers, speech recordings remain sensitive and potentially identifiable biometric data. We therefore treated the recordings and participant-level labels as protected research data. Access was limited to the research workflow, and we made no attempt to identify speakers, recover personal information, or link participants to external records. The ALM and PSE models were run locally; raw audio and participant-level data were not submitted to third-party hosted model services. Generated captions and SC-Para representations were used only for the analyses described in this study.

\paragraph{Human Validation}
The small-scale validation described in Appendix~\ref{app:human_validation} was conducted internally by two graduate researchers. No external annotators were recruited and no additional compensation was provided. Annotators evaluated only whether the generated captions and constituent paralinguistic states were consistent with perceived vocal evidence; they were not asked to infer participant identity, life circumstances, mental state, or clinical diagnosis.

\paragraph{Intended Use and Clinical Interpretation}
ParaCalib is a research framework for studying context-sensitive paralinguistic representations. It is not a medical device and should not be used for diagnosis, triage, treatment decisions, eligibility determinations, or covert monitoring. The prediction targets in this study are binary labels derived from PHQ-8 and PHQ-9 thresholds. These labels indicate questionnaire-defined symptom burden and are not equivalent to independently established clinical diagnoses. Any prospective clinical use would require external validation, appropriate regulatory and ethical review, and qualified human oversight.

\paragraph{Bias and Generalizability}
Speech-based prediction can be affected by language, accent, dialect, age, gender, culture, recording conditions, and access to care. DAIC-WOZ and MODMA represent limited English- and Mandarin-speaking cohorts and were evaluated independently; the experiments do not establish cross-lingual or cross-population generalizability. ALM-generated descriptions and PSE-derived states may also reflect biases or errors inherited from their pretrained models. Accordingly, the extracted states should be interpreted as model-derived representations rather than clinically observed symptoms. Evaluation across larger and more diverse populations is necessary before considering broader use.

\section*{Acknowledgments}
This work was supported by the Lien Foundation, Singapore.

\bibliography{custom}

@misc{who2025depression,
  title        = {Depressive disorder (depression)},
  author       = {{World Health Organization}},
  year         = {2025},
  note         = {WHO Fact Sheet, accessed 2026}
}

@article{cummins2015review,
  title   = {A review of depression and suicide risk assessment using speech analysis},
  author  = {Cummins, Nicholas and Scherer, Stefan and Krajewski, Jarek and Schnieder, Sebastian and Epps, Julien and Quatieri, Thomas F.},
  journal = {Speech Communication},
  volume  = {71},
  pages   = {10--49},
  year    = {2015}
}

@book{dsm5tr2022,
    author = {{American Psychiatric Association}},
    title = {Diagnostic and statistical manual of mental disorders},
    edition = {5th ed., text rev.},
    year = {2022},
    publisher = {American Psychiatric Association},
    doi = {10.1176/appi.books.9780890425787},
    url = {https://doi.org/10.1176/appi.books.9780890425787}
}

@article{li2026depflow,
  title={DepFlow: Disentangled Speech Generation to Mitigate Semantic Bias in Depression Detection},
  author={Li, Yuxin and Zhang, Xiangyu and Li, Yifei and Guo, Zhiwei and Zhang, Haoyang and Chng, Eng Siong and Guan, Cuntai},
  journal={arXiv preprint arXiv:2601.00303},
  year={2026}
}

@article{snyder2015musan,
  title={Musan: A music, speech, and noise corpus},
  author={Snyder, David and Chen, Guoguo and Povey, Daniel},
  journal={arXiv preprint arXiv:1510.08484},
  year={2015}
}

@article{eyben2016geneva,
  title   = {The Geneva Minimalistic Acoustic Parameter Set ({GeMAPS}) for Voice Research and Affective Computing},
  author  = {Eyben, Florian and Scherer, Klaus R. and Schuller, Bj{\"o}rn W. and Sundberg, Johan and Andr{\'e}, Elisabeth and Busso, Carlos and Devillers, Laurence Y. and Epps, Julien and Laukka, Petri and Narayanan, Shrikanth S. and Truong, Khiet P.},
  journal = {IEEE Transactions on Affective Computing},
  volume  = {7},
  number  = {2},
  pages   = {190--202},
  year    = {2016},
  doi     = {10.1109/TAFFC.2015.2457417}
}

@misc{xu2025qwen3omnitechnicalreport,
      title={Qwen3-Omni Technical Report}, 
      author={Jin Xu and Zhifang Guo and Hangrui Hu and Yunfei Chu and Xiong Wang and Jinzheng He and Yuxuan Wang and Xian Shi and Ting He and Xinfa Zhu and Yuanjun Lv and Yongqi Wang and Dake Guo and He Wang and Linhan Ma and Pei Zhang and Xinyu Zhang and Hongkun Hao and Zishan Guo and Baosong Yang and Bin Zhang and Ziyang Ma and Xipin Wei and Shuai Bai and Keqin Chen and Xuejing Liu and Peng Wang and Mingkun Yang and Dayiheng Liu and Xingzhang Ren and Bo Zheng and Rui Men and Fan Zhou and Bowen Yu and Jianxin Yang and Le Yu and Jingren Zhou and Junyang Lin},
      year={2025},
      eprint={2509.17765},
      archivePrefix={arXiv},
      primaryClass={cs.CL},
      url={https://arxiv.org/abs/2509.17765}, 
}

@article{Guo_2025,
   title={DeepSeek-R1 incentivizes reasoning in LLMs through reinforcement learning},
   volume={645},
   ISSN={1476-4687},
   url={http://dx.doi.org/10.1038/s41586-025-09422-z},
   DOI={10.1038/s41586-025-09422-z},
   number={8081},
   journal={Nature},
   publisher={Springer Science and Business Media LLC},
   author={Guo, Daya and Yang, Dejian and Zhang, Haowei and Song, Junxiao and Wang, Peiyi and Zhu, Qihao and Xu, Runxin and Zhang, Ruoyu and Ma, Shirong and Bi, Xiao and Zhang, Xiaokang and Yu, Xingkai and Wu, Yu and Wu, Z. F. and Gou, Zhibin and Shao, Zhihong and Li, Zhuoshu and Gao, Ziyi and Liu, Aixin and Xue, Bing and Wang, Bingxuan and Wu, Bochao and Feng, Bei and Lu, Chengda and Zhao, Chenggang and Deng, Chengqi and Ruan, Chong and Dai, Damai and Chen, Deli and Ji, Dongjie and Li, Erhang and Lin, Fangyun and Dai, Fucong and Luo, Fuli and Hao, Guangbo and Chen, Guanting and Li, Guowei and Zhang, H. and Xu, Hanwei and Ding, Honghui and Gao, Huazuo and Qu, Hui and Li, Hui and Guo, Jianzhong and Li, Jiashi and Chen, Jingchang and Yuan, Jingyang and Tu, Jinhao and Qiu, Junjie and Li, Junlong and Cai, J. L. and Ni, Jiaqi and Liang, Jian and Chen, Jin and Dong, Kai and Hu, Kai and You, Kaichao and Gao, Kaige and Guan, Kang and Huang, Kexin and Yu, Kuai and Wang, Lean and Zhang, Lecong and Zhao, Liang and Wang, Litong and Zhang, Liyue and Xu, Lei and Xia, Leyi and Zhang, Mingchuan and Zhang, Minghua and Tang, Minghui and Zhou, Mingxu and Li, Meng and Wang, Miaojun and Li, Mingming and Tian, Ning and Huang, Panpan and Zhang, Peng and Wang, Qiancheng and Chen, Qinyu and Du, Qiushi and Ge, Ruiqi and Zhang, Ruisong and Pan, Ruizhe and Wang, Runji and Chen, R. J. and Jin, R. L. and Chen, Ruyi and Lu, Shanghao and Zhou, Shangyan and Chen, Shanhuang and Ye, Shengfeng and Wang, Shiyu and Yu, Shuiping and Zhou, Shunfeng and Pan, Shuting and Li, S. S. and Zhou, Shuang and Wu, Shaoqing and Yun, Tao and Pei, Tian and Sun, Tianyu and Wang, T. and Zeng, Wangding and Liu, Wen and Liang, Wenfeng and Gao, Wenjun and Yu, Wenqin and Zhang, Wentao and Xiao, W. L. and An, Wei and Liu, Xiaodong and Wang, Xiaohan and Chen, Xiaokang and Nie, Xiaotao and Cheng, Xin and Liu, Xin and Xie, Xin and Liu, Xingchao and Yang, Xinyu and Li, Xinyuan and Su, Xuecheng and Lin, Xuheng and Li, X. Q. and Jin, Xiangyue and Shen, Xiaojin and Chen, Xiaosha and Sun, Xiaowen and Wang, Xiaoxiang and Song, Xinnan and Zhou, Xinyi and Wang, Xianzu and Shan, Xinxia and Li, Y. K. and Wang, Y. Q. and Wei, Y. X. and Zhang, Yang and Xu, Yanhong and Li, Yao and Zhao, Yao and Sun, Yaofeng and Wang, Yaohui and Yu, Yi and Zhang, Yichao and Shi, Yifan and Xiong, Yiliang and He, Ying and Piao, Yishi and Wang, Yisong and Tan, Yixuan and Ma, Yiyang and Liu, Yiyuan and Guo, Yongqiang and Ou, Yuan and Wang, Yuduan and Gong, Yue and Zou, Yuheng and He, Yujia and Xiong, Yunfan and Luo, Yuxiang and You, Yuxiang and Liu, Yuxuan and Zhou, Yuyang and Zhu, Y. X. and Huang, Yanping and Li, Yaohui and Zheng, Yi and Zhu, Yuchen and Ma, Yunxian and Tang, Ying and Zha, Yukun and Yan, Yuting and Ren, Z. Z. and Ren, Zehui and Sha, Zhangli and Fu, Zhe and Xu, Zhean and Xie, Zhenda and Zhang, Zhengyan and Hao, Zhewen and Ma, Zhicheng and Yan, Zhigang and Wu, Zhiyu and Gu, Zihui and Zhu, Zijia and Liu, Zijun and Li, Zilin and Xie, Ziwei and Song, Ziyang and Pan, Zizheng and Huang, Zhen and Xu, Zhipeng and Zhang, Zhongyu and Zhang, Zhen},
   year={2025},
   month = sep,
   pages = {633--638} }

@misc{yang2025qwen3technicalreport,
      title={Qwen3 Technical Report}, 
      author={An Yang and Anfeng Li and Baosong Yang and Beichen Zhang and Binyuan Hui and Bo Zheng and Bowen Yu and Chang Gao and Chengen Huang and Chenxu Lv and Chujie Zheng and Dayiheng Liu and Fan Zhou and Fei Huang and Feng Hu and Hao Ge and Haoran Wei and Huan Lin and Jialong Tang and Jian Yang and Jianhong Tu and Jianwei Zhang and Jianxin Yang and Jiaxi Yang and Jing Zhou and Jingren Zhou and Junyang Lin and Kai Dang and Keqin Bao and Kexin Yang and Le Yu and Lianghao Deng and Mei Li and Mingfeng Xue and Mingze Li and Pei Zhang and Peng Wang and Qin Zhu and Rui Men and Ruize Gao and Shixuan Liu and Shuang Luo and Tianhao Li and Tianyi Tang and Wenbiao Yin and Xingzhang Ren and Xinyu Wang and Xinyu Zhang and Xuancheng Ren and Yang Fan and Yang Su and Yichang Zhang and Yinger Zhang and Yu Wan and Yuqiong Liu and Zekun Wang and Zeyu Cui and Zhenru Zhang and Zhipeng Zhou and Zihan Qiu},
      year={2025},
      eprint={2505.09388},
      archivePrefix={arXiv},
      primaryClass={cs.CL},
      url={https://arxiv.org/abs/2505.09388}, 
}

@inproceedings{reimers2019sentencebert,
  title     = {Sentence-{BERT}: Sentence Embeddings using {S}iamese {BERT}-Networks},
  author    = {Reimers, Nils and Gurevych, Iryna},
  booktitle = {Proceedings of the 2019 Conference on Empirical Methods in Natural Language Processing and the 9th International Joint Conference on Natural Language Processing},
  pages     = {3982--3992},
  publisher = {Association for Computational Linguistics},
  year      = {2019},
  doi       = {10.18653/v1/D19-1410},
  url       = {https://aclanthology.org/D19-1410/}
}

@inproceedings{kwon2023vllm,
  title     = {Efficient Memory Management for Large Language Model Serving with {PagedAttention}},
  author    = {Kwon, Woosuk and Li, Zhuohan and Zhuang, Siyuan and Sheng, Ying and Zheng, Lianmin and Yu, Cody Hao and Gonzalez, Joseph E. and Zhang, Hao and Stoica, Ion},
  booktitle = {Proceedings of the 29th Symposium on Operating Systems Principles},
  pages     = {611--626},
  publisher = {Association for Computing Machinery},
  year      = {2023},
  doi       = {10.1145/3600006.3613165}
}

@misc{tian2025stepaudior1technicalreport,
      title={Step-Audio-R1 Technical Report}, 
      author={Fei Tian and Xiangyu Tony Zhang and Yuxin Zhang and Haoyang Zhang and Yuxin Li and Daijiao Liu and Yayue Deng and Donghang Wu and Jun Chen and Liang Zhao and Chengyuan Yao and Hexin Liu and Eng Siong Chng and Xuerui Yang and Xiangyu Zhang and Daxin Jiang and Gang Yu},
      year={2025},
      eprint={2511.15848},
      archivePrefix={arXiv},
      primaryClass={cs.AI},
      url={https://arxiv.org/abs/2511.15848}, 
}

@article{zhang2025speecht,
  title={SpeechT-RAG: Reliable Depression Detection in LLMs with Retrieval-Augmented Generation Using Speech Timing Information},
  author={Zhang, Xiangyu and Liu, Hexin and Zhang, Qiquan and Ahmed, Beena and Epps, Julien},
  journal={arXiv preprint arXiv:2502.10950},
  year={2025}
}

@inproceedings{morales2018linguistically,
  title={A linguistically-informed fusion approach for multimodal depression detection},
  author={Morales, Michelle and Scherer, Stefan and Levitan, Rivka},
  booktitle={proceedings of the fifth workshop on computational linguistics and clinical psychology: from keyboard to clinic},
  pages={13--24},
  year={2018}
}

@article{kroenke2001phq,
  title={The PHQ-9: validity of a brief depression severity measure},
  author={Kroenke, Kurt and Spitzer, Robert L and Williams, Janet BW},
  journal={Journal of general internal medicine},
  volume={16},
  number={9},
  pages={606--613},
  year={2001},
  publisher={Wiley Online Library}
}

@article{kroenke2009phq,
  title={The PHQ-8 as a measure of current depression in the general population},
  author={Kroenke, Kurt and Strine, Tara W and Spitzer, Robert L and Williams, Janet BW and Berry, Joyce T and Mokdad, Ali H},
  journal={Journal of affective disorders},
  volume={114},
  number={1-3},
  pages={163--173},
  year={2009},
  publisher={Elsevier}
}

@article{chandra2025emotionrankclap,
  title={EmotionRankCLAP: Bridging Natural Language Speaking Styles and Ordinal Speech Emotion via Rank-N-Contrast},
  author={Chandra, Shreeram Suresh and Goncalves, Lucas and Lu, Junchen and Busso, Carlos and Sisman, Berrak},
  journal={arXiv preprint arXiv:2505.23732},
  year={2025}
}

@article{alpert2001reflections,
  title={Reflections of depression in acoustic measures of the patient’s speech},
  author={Alpert, Murray and Pouget, Enrique R and Silva, Raul R},
  journal={Journal of affective disorders},
  volume={66},
  number={1},
  pages={59--69},
  year={2001},
  publisher={Elsevier}
}

@article{zhao2025hears,
  title={It hears, it sees too: Multi-modal LLM for depression detection by integrating visual understanding into audio language models},
  author={Zhao, Xiangyu and Shen, Yaling and Jiang, Yiwen and Wang, Zimu and Liu, Jiahe and Cheng, Maxmartwell H and Oliveira, Guilherme C and Desimone, Robert and Dwyer, Dominic and Ge, Zongyuan},
  journal={arXiv preprint arXiv:2511.19877},
  year={2025}
}

@inproceedings{dubagunta2019learning,
  title={Learning voice source related information for depression detection},
  author={Dubagunta, S Pavankumar and Vlasenko, Bogdan and Doss, Mathew Magimai-},
  booktitle={ICASSP 2019-2019 IEEE International Conference on Acoustics, Speech and Signal Processing (ICASSP)},
  pages={6525--6529},
  year={2019},
  organization={IEEE}
}

@inproceedings{zhang2024llms,
  title={When llms meets acoustic landmarks: An efficient approach to integrate speech into large language models for depression detection},
  author={Zhang, Xiangyu and Liu, Hexin and Xu, Kaishuai and Zhang, Qiquan and Liu, Daijiao and Ahmed, Beena and Epps, Julien},
  booktitle={Proceedings of the 2024 Conference on Empirical Methods in Natural Language Processing},
  pages={146--158},
  year={2024}
}

@inproceedings{zhang-etal-2025-beyond-classification,
    title = "Beyond Classification: Towards Speech Emotion Reasoning with Multitask {A}udio{LLM}s",
    author = "Zhang, Wenyu  and
      He, Yingxu  and
      Lin, Geyu  and
      Liu, Zhuohan  and
      Sun, Shuo  and
      Wang, Bin  and
      Zou, Xunlong  and
      Wong, Jeremy H. M.  and
      Wang, Qiongqiong  and
      Sailor, Hardik Bhupendra  and
      Chen, Nancy F.  and
      Aw, AiTi",
    editor = "Inui, Kentaro  and
      Sakti, Sakriani  and
      Wang, Haofen  and
      Wong, Derek F.  and
      Bhattacharyya, Pushpak  and
      Banerjee, Biplab  and
      Ekbal, Asif  and
      Chakraborty, Tanmoy  and
      Singh, Dhirendra Pratap",
    booktitle = "Proceedings of the 14th International Joint Conference on Natural Language Processing and the 4th Conference of the Asia-Pacific Chapter of the Association for Computational Linguistics",
    month = dec,
    year = "2025",
    address = "Mumbai, India",
    publisher = "The Asian Federation of Natural Language Processing and The Association for Computational Linguistics",
    url = "https://aclanthology.org/2025.ijcnlp-long.62/",
    doi = "10.18653/v1/2025.ijcnlp-long.62",
    pages = "1132--1148",
    ISBN = "979-8-89176-298-5"
}

@inproceedings{bailey2021gender,
  title={Gender bias in depression detection using audio features},
  author={Bailey, Andrew and Plumbley, Mark D},
  booktitle={2021 29th European Signal Processing Conference (EUSIPCO)},
  pages={596--600},
  year={2021},
  organization={IEEE}
}

@article{koops2023speech,
  title={Speech as a biomarker for depression},
  author={Koops, Sanne and Brederoo, Sanne G and De Boer, Janna N and Nadema, Femke G and Voppel, Alban E and Sommer, Iris E},
  journal={CNS \& Neurological Disorders-Drug Targets-CNS \& Neurological Disorders)},
  volume={22},
  number={2},
  pages={152--160},
  year={2023},
  publisher={Bentham Science Publishers direct}
}

@inproceedings{ghosh2024gama,
  title={Gama: A large audio-language model with advanced audio understanding and complex reasoning abilities},
  author={Ghosh, Sreyan and Kumar, Sonal and Seth, Ashish and Evuru, Chandra Kiran Reddy and Tyagi, Utkarsh and Sakshi, S and Nieto, Oriol and Duraiswami, Ramani and Manocha, Dinesh},
  booktitle={Proceedings of the 2024 Conference on Empirical Methods in Natural Language Processing},
  pages={6288--6313},
  year={2024}
}

@article{Cai_2022,
   title={A multi-modal open dataset for mental-disorder analysis},
   volume={9},
   ISSN={2052-4463},
   url={http://dx.doi.org/10.1038/s41597-022-01211-x},
   DOI={10.1038/s41597-022-01211-x},
   number={1},
   journal={Scientific Data},
   publisher={Springer Science and Business Media LLC},
   author={Cai, Hanshu and Yuan, Zhenqin and Gao, Yiwen and Sun, Shuting and Li, Na and Tian, Fuze and Xiao, Han and Li, Jianxiu and Yang, Zhengwu and Li, Xiaowei and Zhao, Qinglin and Liu, Zhenyu and Yao, Zhijun and Yang, Minqiang and Peng, Hong and Zhu, Jing and Zhang, Xiaowei and Gao, Guoping and Zheng, Fang and Li, Rui and Guo, Zhihua and Ma, Rong and Yang, Jing and Zhang, Lan and Hu, Xiping and Li, Yumin and Hu, Bin},
   year={2022},
   month=Apr }

@inproceedings{gratch-etal-2014-distress,
    title = "The Distress Analysis Interview Corpus of human and computer interviews",
    author = "Gratch, Jonathan  and
      Artstein, Ron  and
      Lucas, Gale  and
      Stratou, Giota  and
      Scherer, Stefan  and
      Nazarian, Angela  and
      Wood, Rachel  and
      Boberg, Jill  and
      DeVault, David  and
      Marsella, Stacy  and
      Traum, David  and
      Rizzo, Skip  and
      Morency, Louis-Philippe",
    editor = "Calzolari, Nicoletta  and
      Choukri, Khalid  and
      Declerck, Thierry  and
      Loftsson, Hrafn  and
      Maegaard, Bente  and
      Mariani, Joseph  and
      Moreno, Asuncion  and
      Odijk, Jan  and
      Piperidis, Stelios",
    booktitle = "Proceedings of the Ninth International Conference on Language Resources and Evaluation ({LREC}'14)",
    month = may,
    year = "2014",
    address = "Reykjavik, Iceland",
    publisher = "European Language Resources Association (ELRA)",
    url = "https://aclanthology.org/L14-1421/",
    pages = "3123--3128"
}

@misc{chu2023qwenaudioadvancinguniversalaudio,
      title={Qwen-Audio: Advancing Universal Audio Understanding via Unified Large-Scale Audio-Language Models}, 
      author={Yunfei Chu and Jin Xu and Xiaohuan Zhou and Qian Yang and Shiliang Zhang and Zhijie Yan and Chang Zhou and Jingren Zhou},
      year={2023},
      eprint={2311.07919},
      archivePrefix={arXiv},
      primaryClass={eess.AS},
      url={https://arxiv.org/abs/2311.07919}, 
}

@misc{chu2024qwen2audiotechnicalreport,
      title={Qwen2-Audio Technical Report}, 
      author={Yunfei Chu and Jin Xu and Qian Yang and Haojie Wei and Xipin Wei and Zhifang Guo and Yichong Leng and Yuanjun Lv and Jinzheng He and Junyang Lin and Chang Zhou and Jingren Zhou},
      year={2024},
      eprint={2407.10759},
      archivePrefix={arXiv},
      primaryClass={eess.AS},
      url={https://arxiv.org/abs/2407.10759}, 
}

@inproceedings{tang2024salmonn,
  title={Salmonn: Towards generic hearing abilities for large language models},
  author={Tang, Changli and Yu, Wenyi and Sun, Guangzhi and Chen, Xianzhao and Tan, Tian and Li, Wei and Lu, Lu and Ma, Zejun and Zhang, Chao},
  booktitle={International Conference on Learning Representations},
  volume={2024},
  pages={16607--16629},
  year={2024}
}

@inproceedings{xu2024secap,
  title={Secap: Speech emotion captioning with large language model},
  author={Xu, Yaoxun and Chen, Hangting and Yu, Jianwei and Huang, Qiaochu and Wu, Zhiyong and Zhang, Shi-Xiong and Li, Guangzhi and Luo, Yi and Gu, Rongzhi},
  booktitle={Proceedings of the AAAI Conference on Artificial Intelligence},
  volume={38},
  pages={19323--19331},
  year={2024}
}

@inproceedings{ma2016depaudionet,
  title={Depaudionet: An efficient deep model for audio based depression classification},
  author={Ma, Xingchen and Yang, Hongyu and Chen, Qiang and Huang, Di and Wang, Yunhong},
  booktitle={Proceedings of the 6th international workshop on audio/visual emotion challenge},
  pages={35--42},
  year={2016}
}

@inproceedings{valstar2013avec,
  title={Avec 2013: the continuous audio/visual emotion and depression recognition challenge},
  author={Valstar, Michel and Schuller, Bj{\"o}rn and Smith, Kirsty and Eyben, Florian and Jiang, Bihan and Bilakhia, Sanjay and Schnieder, Sebastian and Cowie, Roddy and Pantic, Maja},
  booktitle={Proceedings of the 3rd ACM international workshop on Audio/visual emotion challenge},
  pages={3--10},
  year={2013}
}

@inproceedings{Othmani2021,
  title={Towards robust deep neural networks for affect and depression recognition from speech},
  author={Othmani, Alice and Kadoch, Daoud and Bentounes, Kamil and Rejaibi, Emna and Alfred, Romain and Hadid, Abdenour},
  booktitle={Pattern Recognition. ICPR International Workshops and Challenges: Virtual Event, January 10--15, 2021, Proceedings, Part II},
  pages={5--19},
  year={2021},
  organization={Springer}
}

@inproceedings{radford2023robust,
  title={Robust speech recognition via large-scale weak supervision},
  author={Radford, Alec and Kim, Jong Wook and Xu, Tao and Brockman, Greg and McLeavey, Christine and Sutskever, Ilya},
  booktitle={International conference on machine learning},
  pages={28492--28518},
  year={2023},
  organization={PMLR}
}

@inproceedings{10.1145/3347320.3357688,
author = {Ringeval, Fabien and Schuller, Bj{\"o}rn and Valstar, Michel and Cummins, Nicholas and Cowie, Roddy and Tavabi, Leili and Schmitt, Maximilian and Alisamir, Sina and Amiriparian, Shahin and Messner, Eva-Maria and Song, Siyang and Liu, Shuo and Zhao, Ziping and Mallol-Ragolta, Adria and Ren, Zhao and Soleymani, Mohammad and Pantic, Maja},
title = {AVEC 2019 Workshop and Challenge: State-of-Mind, Detecting Depression with AI, and Cross-Cultural Affect Recognition},
year = {2019},
isbn = {9781450369138},
publisher = {Association for Computing Machinery},
address = {New York, NY, USA},
url = {https://doi.org/10.1145/3347320.3357688},
doi = {10.1145/3347320.3357688},
booktitle = {Proceedings of the 9th International on Audio/Visual Emotion Challenge and Workshop},
pages = {3–12},
numpages = {10},
location = {Nice, France},
series = {AVEC '19}
}

@article{chlasta2019automated,
  title={Automated speech-based screening of depression using deep convolutional neural networks},
  author={Chlasta, Karol and Wo{\l}k, Krzysztof and Krejtz, Izabela},
  journal={Procedia Computer Science},
  volume={164},
  pages={618--628},
  year={2019},
  publisher={Elsevier}
}

@inproceedings{wang2022ecapa,
  title={ECAPA-TDNN Based Depression Detection from Clinical Speech.},
  author={Wang, Dong and Ding, Yanhui and Zhao, Qing and Yang, Peilin and Tan, Shuping and Li, Ya},
  booktitle={Interspeech},
  pages={3333--3337},
  year={2022}
}

@article{Salekin2018,
  title={A weakly supervised learning framework for detecting social anxiety and depression},
  author={Salekin, Asif and Eberle, Jeremy W and Glenn, Jeffrey J and Teachman, Bethany A and Stankovic, John A},
  journal={Proceedings of the ACM on interactive, mobile, wearable and ubiquitous technologies},
  volume={2},
  number={2},
  pages={1--26},
  year={2018},
  publisher={ACM New York, NY, USA}
}

@inproceedings{zhao2020hierarchical,
  title={Hierarchical attention transfer networks for depression assessment from speech},
  author={Zhao, Ziping and Bao, Zhongtian and Zhang, Zixing and Cummins, Nicholas and Wang, Haishuai and Schuller, Bj{\"o}rn},
  booktitle={ICASSP 2020-2020 IEEE international conference on acoustics, speech and signal processing (ICASSP)},
  pages={7159--7163},
  year={2020},
  organization={IEEE}
}

@article{muzammel2020audvowelconsnet,
  title={AudVowelConsNet: A phoneme-level based deep CNN architecture for clinical depression diagnosis},
  author={Muzammel, Muhammad and Salam, Hanan and Hoffmann, Yann and Chetouani, Mohamed and Othmani, Alice},
  journal={Machine Learning with Applications},
  volume={2},
  pages={100005},
  year={2020},
  publisher={Elsevier}
}

@article{vazquez2020automatic,
  title={Automatic detection of depression in speech using ensemble convolutional neural networks},
  author={V{\'a}zquez-Romero, Adri{\'a}n and Gallardo-Antol{\'\i}n, Ascensi{\'o}n},
  journal={Entropy},
  volume={22},
  number={6},
  pages={688},
  year={2020},
  publisher={MDPI}
}

@inproceedings{saidi2020hybrid,
  title={Hybrid CNN-SVM classifier for efficient depression detection system},
  author={Saidi, Afef and Othman, Slim Ben and Saoud, Slim Ben},
  booktitle={2020 4th International Conference on Advanced Systems and Emergent Technologies (IC\_ASET)},
  pages={229--234},
  year={2020},
  organization={IEEE}
}

@article{li2025hierarchical,
  title={Hierarchical Self-Supervised Representation Learning for Depression Detection from Speech},
  author={Li, Yuxin and Chng, Eng Siong and Guan, Cuntai},
  journal={arXiv preprint arXiv:2510.08593},
  year={2025}
}

@article{li2025automated,
  title={Automated depression detection from text and audio: A systematic review},
  author={Li, Yuxin and Kumbale, Sinchana and Chen, Yanru and Surana, Tanmay and Chng, Eng Siong and Guan, Cuntai},
  journal={IEEE Journal of Biomedical and Health Informatics},
  year={2025},
  publisher={IEEE}
}

@article{ellgring1996vocal,
  title   = {Vocal indicators of mood change in depression},
  author  = {Ellgring, Heiner and Scherer, Klaus R.},
  journal = {Journal of Nonverbal Behavior},
  volume  = {20},
  number  = {2},
  pages   = {83--110},
  year    = {1996}
}

@article{yamamoto2020speech,
  title   = {Using speech recognition technology to investigate the association between timing-related speech features and depression severity},
  author  = {Yamamoto, Masahiro and Takamiya, Akihiro and Sawada, Kazutaka and Yoshimura, Masaru and Kitazawa, Masahiro and Liang, Kuang and Fujita, Takashi and Mimura, Masaru and Kishimoto, Taishiro},
  journal = {PLOS ONE},
  volume  = {15},
  number  = {9},
  pages   = {e0238726},
  year    = {2020}
}

@inproceedings{Zhang2021Depa,
  title={Depa: Self-supervised audio embedding for depression detection},
  author={Zhang, Pingyue and Wu, Mengyue and Dinkel, Heinrich and Yu, Kai},
  booktitle={Proceedings of the 29th ACM international conference on multimedia},
  pages={135--143},
  year={2021}
}

@inproceedings{wu2023self,
  title={Self-supervised representations in speech-based depression detection},
  author={Wu, Wen and Zhang, Chao and Woodland, Philip C},
  booktitle={ICASSP 2023-2023 IEEE International Conference on Acoustics, Speech and Signal Processing (ICASSP)},
  pages={1--5},
  year={2023},
  organization={IEEE}
}

@inproceedings{ravi2022step,
  title={A step towards preserving speakers’ identity while detecting depression via speaker disentanglement},
  author={Ravi, Vijay and Wang, Jinhan and Flint, Jonathan and Alwan, Abeer},
  booktitle={Interspeech},
  volume={2022},
  pages={3338},
  year={2022}
}

@inproceedings{Toto2021,
  title={Audibert: A deep transfer learning multimodal classification framework for depression screening},
  author={Toto, Ermal and Tlachac, ML and Rundensteiner, Elke A},
  booktitle={Proceedings of the 30th ACM international conference on information \& knowledge management},
  pages={4145--4154},
  year={2021}
}

@inproceedings{baevski2020wav2vec,
  title     = {wav2vec 2.0: A framework for self-supervised learning of speech representations},
  author    = {Baevski, Alexei and Zhou, Henry and Mohamed, Abdelrahman and Auli, Michael},
  booktitle = {Advances in Neural Information Processing Systems},
  volume    = {33},
  pages     = {12449--12460},
  year      = {2020}
}

@article{hsu2021hubert,
  title   = {{HuBERT}: Self-supervised speech representation learning by masked prediction of hidden units},
  author  = {Hsu, Wei-Ning and Bolte, Benjamin and Tsai, Yao-Hung Hubert and Lakhotia, Kushal and Salakhutdinov, Ruslan and Mohamed, Abdelrahman},
  journal = {IEEE/ACM Transactions on Audio, Speech, and Language Processing},
  volume  = {29},
  pages   = {3451--3460},
  year    = {2021}
}

@article{Chen_2022,
   title={WavLM: Large-Scale Self-Supervised Pre-Training for Full Stack Speech Processing},
   volume={16},
   ISSN={1941-0484},
   url={http://dx.doi.org/10.1109/JSTSP.2022.3188113},
   DOI={10.1109/jstsp.2022.3188113},
   number={6},
   journal={IEEE Journal of Selected Topics in Signal Processing},
   publisher={Institute of Electrical and Electronics Engineers (IEEE)},
   author={Chen, Sanyuan and Wang, Chengyi and Chen, Zhengyang and Wu, Yu and Liu, Shujie and Chen, Zhuo and Li, Jinyu and Kanda, Naoyuki and Yoshioka, Takuya and Xiao, Xiong and Wu, Jian and Zhou, Long and Ren, Shuo and Qian, Yanmin and Qian, Yao and Wu, Jian and Zeng, Michael and Yu, Xiangzhan and Wei, Furu},
   year={2022},
   month=Oct, pages={1505–1518} }

@article{cole2015prosody,
  title   = {Prosody in context: A review},
  author  = {Cole, Jennifer},
  journal = {Language, Cognition and Neuroscience},
  volume  = {30},
  number  = {1-2},
  pages   = {1--31},
  year    = {2015}
}

@article{wagner2010prosody,
  title   = {Experimental and theoretical advances in prosody: A review},
  author  = {Wagner, Michael and Watson, Duane G.},
  journal = {Language and Cognitive Processes},
  volume  = {25},
  number  = {7-9},
  pages   = {905--945},
  year    = {2010}
}

@inproceedings{wang2023non,
  title={Non-uniform speaker disentanglement for depression detection from raw speech signals},
  author={Wang, Jinhan and Ravi, Vijay and Alwan, Abeer},
  booktitle={Interspeech},
  volume={2023},
  pages={2343},
  year={2023}
}

@article{chen2022speechformer,
  title={SpeechFormer: A hierarchical efficient framework incorporating the characteristics of speech},
  author={Chen, Weidong and Xing, Xiaofen and Xu, Xiangmin and Pang, Jianxin and Du, Lan},
  journal={arXiv preprint arXiv:2203.03812},
  year={2022}
}

@article{galatzer2023machine,
  title={Machine learning and the digital measurement of psychological health},
  author={Galatzer-Levy, Isaac R and Onnela, Jukka-Pekka},
  journal={Annual Review of Clinical Psychology},
  volume={19},
  pages={133--154},
  year={2023},
  publisher={Annual Reviews}
}

@article{CHEN2026113367,
title = {Towards stable cross-domain depression recognition under missing modalities},
journal = {Pattern Recognition},
volume = {177},
pages = {113367},
year = {2026},
issn = {0031-3203},
doi = {https://doi.org/10.1016/j.patcog.2026.113367},
url = {https://www.sciencedirect.com/science/article/pii/S0031320326003328},
author = {Jiuyi Chen and Mingkui Tan and Haifeng Lu and Qiuna Xu and Zhihua Wang and Runhao Zeng and Xiping Hu}
}

\appendix

\section{Experimental Setup \& Implementation Details} \label{app:setup}

\subsection{Dataset Details and Preprocessing}
We correct two known annotation inconsistencies. For DAIC-WOZ, we relabel participant 409 as depressed following Bailey et al.~\citep{bailey2021gender}. For MODMA, we use PHQ-based binary labels rather than the original clinician-defined grouping and relabel participant 2010037 as non-depression based on the PHQ score, following Li et al.~\citep{li2025hierarchical}.

For DAIC-WOZ, raw audio is segmented into participant utterances using transcript timestamps. We correct transcript--audio misalignment by shifting timestamps by 34.000~s for participant 318, 3.355~s for participant 321, 6.070~s for participant 341, and 16.540~s for participant 362. MODMA is distributed with sentence-level segmentation.

\subsection{Training Details, Architectures, and Splits}
All compared classifiers are trained under the same optimization, batching, device, and early-stopping settings on a single NVIDIA RTX 6000 GPU.

\begin{table}[t]
\centering

\resizebox{\columnwidth}{!}{%
\begin{tabular}{ll}
\toprule
Category & Hyperparameter \& Value \\
\midrule
Input & Unit: SC-Para vector (Dim 7) \\
& Grouping: Valid vectors per participant \\
\midrule
Attention & Projection: Linear, $7 \rightarrow 64$, Tanh \\
& Query: Learned 64-d vector \\
& Weights: Softmax over utterances \\
& Readout: Weighted sum of 7-d vectors \\
\midrule
Classifier & Architecture: 2-layer MLP (64, 32) \\
& Norm: BatchNorm (after 1st layer) \\
& Act./Dropout: ReLU / 0.3 \\
\midrule
Optim. & Optimizer: AdamW ($5\times10^{-4}$ / $1\times10^{-4}$) \\
& Pos. Weight: $N_{\mathrm{neg}}/N_{\mathrm{pos}}$, Clip: 1.0 \\
& LR scheduler: ReduceLROnPlateau \\
\midrule
Training & Max epochs/batch: 50 / 4 \\
& Early stopping patience: 10 \\
\bottomrule
\end{tabular}
}
\caption{Implementation and training configuration.}
\label{tab:hyperparams}

\vspace{1em} 

\small
\resizebox{\columnwidth}{!}{
\begin{tabular}{lrrr}
\toprule
Comparison & $\Delta$ Macro-F1 & 95\% CI & $p_{\mathrm{adj}}$ \\
\midrule
\multicolumn{4}{c}{\textit{SC-Para vs. HAREN-CTC}} \\
DAIC-WOZ & 14.03 & [6.76, 21.30] & .0232 \\
MODMA & 17.46 & [3.68, 31.24] & .0734 \\
\midrule
\multicolumn{4}{c}{\textit{SC-Para vs. Caption Embedding}} \\
DAIC-WOZ & 5.37 & [-9.18, 19.92] & .7269 \\
MODMA & 4.80 & [-15.23, 24.84] & .7269 \\
\bottomrule
\end{tabular}
}
\caption{Paired Macro-F1 comparisons across the five matched evaluation units. Confidence intervals are unadjusted 95\% paired-difference intervals; $p$-values are adjusted using the Holm procedure across the four comparisons.}
\label{tab:sig}
\end{table}

\paragraph{Architecture Details:}
The MLP baseline applies a two-hidden-layer classifier ($\sim$2.7K params). The LSTM applies a two-layer LSTM ($\sim$58K). The Bi-LSTM uses two bidirectional layers ($\sim$147K). The Transformer uses two encoder layers and mean pooling ($\sim$106K). Attentive MIL uses attention pooling followed by the same MLP
($\sim$3.3K parameters in total). A full performance breakdown is shown in Table~\ref{tab:classifiers}.

\paragraph{Significance and Split-wise Performance:}
We compare macro-F1 scores using a two-sided paired $t$-test across the five matched runs described in Section~\ref{sec:data_evaluation} (Table~\ref{tab:sig}). These tests characterize matched run- or fold-level variation and are not participant-resampling analyses. Table~\ref{tab:splits} reports the training, development, and test performance under the evaluated splits.

\begin{table*}[t]
\centering
\small
\begin{tabular}{llcccc}
\toprule
Dataset & Classifier & Macro-F1 & Sensitivity & Specificity & AUC \\
\midrule
\multirow{6}{*}{DAIC-WOZ}
& Bi-LSTM & 49.6 (3.3) & 40.0 (13.0) & 61.2 (10.4) & 56.8 (3.8) \\
& LSTM & 42.2 (5.0) & 18.6 (14.8) & 70.9 (24.5) & 45.7 (3.7) \\
& Transformer w/ Pos. Enc. & 40.2 (10.2) & 22.9 (43.3) & \textbf{79.4} (44.4) & 55.8 (3.6) \\
& Transformer w/o Pos. Enc. & 60.6 (5.5) & \textbf{85.7} (8.7) & 50.9 (11.0) & 73.2 (3.7) \\
& MLP & 65.5 (5.0) & 58.6 (9.3) & 74.5 (2.7) & 73.7 (4.2) \\
& Attentive MIL & \textbf{71.9} (4.9) & 77.1 (15.5) & 73.3 (6.6) & \textbf{83.3} (2.6) \\
\midrule
\multirow{6}{*}{MODMA}
& LSTM & 79.3 (7.5) & 74.0 (9.4) & 85.3 (14.1) & 94.2 (4.2) \\
& Transformer w/ Pos. Enc. & 77.6 (9.1) & 71.4 (9.4) & 86.4 (16.7) & 94.4 (7.1) \\
& Transformer w/o Pos. Enc. & 81.3 (8.1) & 77.0 (7.3) & 86.0 (14.8) & 95.7 (6.0) \\
& Bi-LSTM & 82.8 (7.5) & 80.6 (6.8) & 85.5 (11.7) & 94.2 (8.1) \\
& MLP & 84.7 (13.1) & 90.0 (10.3) & 81.9 (18.8) & 90.5 (10.9) \\
& Attentive MIL & \textbf{90.5} (11.2) & \textbf{91.0} (12.5) & \textbf{90.0} (14.9) & \textbf{96.4} (6.9) \\
\bottomrule
\end{tabular}
\caption{Classifier comparison using the same utterance-level SC-Para vectors.}
\label{tab:classifiers}

\vspace{1em}

\small
\setlength{\tabcolsep}{4pt}
\begin{tabular}{llccccc}
\toprule
Dataset & Split & Accuracy & Macro-F1 & Sensitivity & Specificity & AUC \\
\midrule
\multirow{3}{*}{DAIC-WOZ}
& Train & 74.9 (3.2) & 70.1 (4.2) & 61.3 (8.7) & 80.3 (2.6) & 78.0 (6.7) \\
& Dev & 74.9 (6.5) & 71.0 (7.6) & 56.7 (12.4) & 84.4 (5.8) & 83.0 (6.9) \\
& Test & 74.5 (4.0) & 71.9 (4.9) & 77.1 (15.5) & 73.3 (6.6) & 83.3 (2.6) \\
\midrule
\multirow{3}{*}{MODMA}
& Train & 92.9 (1.9) & 92.8 (1.9) & 91.5 (3.2) & 94.0 (3.1) & 98.5 (0.7) \\
& Dev & 91.4 (10.5) & 91.1 (10.6) & 88.4 (10.5) & 93.3 (13.1) & 95.8 (7.2) \\
& Test & 90.7 (11.1) & 90.5 (11.2) & 91.0 (12.5) & 90.0 (14.9) & 96.4 (6.9) \\
\bottomrule
\end{tabular}
\caption{Training, development, and test performance of ParaCalib across splits.}
\label{tab:splits}
\end{table*}

\subsection{Model and Reproducibility Settings}\label{app:model_settings}
We use Qwen3-Omni-30B-A3B-Captioner~\citep{xu2025qwen3omnitechnicalreport} and DeepSeek-R1-Distill-Qwen-32B~\citep{Guo_2025} locally with an OpenAI-compatible vLLM backend~\citep{kwon2023vllm} in \texttt{bfloat16} precision. Generation uses temperature $=0$, top-$p=1.0$. Table~\ref{tab:hyperparams} records the downstream optimizer, learning rates, batch size, scheduler, and early-stopping settings; Section~\ref{sec:data_evaluation} specifies the seed and fold aggregation used for evaluation.

The continuous LLD baseline uses 88 utterance-level eGeMAPSv02 functionals~\citep{eyben2016geneva}. The discretized baseline maps five functionals to acoustic proxies (Table~\ref{tab:lld}). For SSL representations, we use the hidden states from the last Transformer layer of Wav2Vec2-large-960h, HuBERT-large-ls960-ft, and WavLM-large. Transcript representations encode utterances with a frozen all-MiniLM-L6-v2~\citep{reimers2019sentencebert}. For Caption Embedding, we encode each complete ALM-generated caption using the same frozen all-MiniLM-L6-v2 encoder and use the resulting 384-dimensional mean-pooled sentence embedding as the utterance-level MIL instance.

\begin{table*}[t]
\centering
\small
\begin{tabular}{lll}
\toprule
SC-Para proxy & eGeMAPSv02 functional & Discretization \\
\midrule
Pitch var. & \texttt{F0semitoneFrom27.5Hz\_sma3nz\_stddevNorm} & Tertiles \\
Speak rate & \texttt{VoicedSegmentsPerSec} & Tertiles \\
Rhythm & \texttt{StddevVoicedSegmentLengthSec} & Median \\
Pauses & \texttt{MeanUnvoicedSegmentLength} & Tertiles \\
Energy & \texttt{loudness\_sma3\_amean} & Tertiles \\
\bottomrule
\end{tabular}

\caption{Mapping used by the discretized LLD baseline.}
\label{tab:lld}
\end{table*}

\section{Additional Ablation and Sensitivity Analyses}
\label{app:ablations}

\subsection{Correlations among Categorical SC-Para Dimensions}

Table~\ref{tab:corr} reports the pairwise Spearman correlations among the five categorical SC-Para dimensions on DAIC-WOZ, and Table~\ref{tab:stab} reports their extraction stability under repeated decoding. Correlation magnitudes reach up to .70, indicating that the dimensions are related but not redundant.

\begin{table}[t]
\centering
\small
\setlength{\tabcolsep}{3pt}
\begin{tabular}{@{}lrrrrr@{}}
\toprule
& Pitch & Rate & Rhythm & Pauses & Energy \\
\midrule
Pitch  & 1.000 &  .586 & -.236 & -.265 &  .674 \\
Rate   &  .586 & 1.000 & -.336 & -.400 &  .703 \\
Rhythm & -.236 & -.336 & 1.000 &  .589 & -.296 \\
Pauses & -.265 & -.400 &  .589 & 1.000 & -.349 \\
Energy &  .674 &  .703 & -.296 & -.349 & 1.000 \\
\bottomrule
\end{tabular}
\caption{Pairwise Spearman correlations among the five categorical SC-Para dimensions on DAIC-WOZ.}
\label{tab:corr}

\vspace{1em}

\small
\setlength{\tabcolsep}{5pt}
\begin{tabular}{@{}lc@{}}
\toprule
SC-Para dimension & Exact agreement \\
\midrule
Pitch variability    & 99 / 100 (99\%) \\
Speaking rate        & 99 / 100 (99\%) \\
Rhythm regularity    & 99 / 100 (99\%) \\
Pause pattern        & 98 / 100 (98\%) \\
Vocal energy         & 98 / 100 (98\%) \\
Depression evidence  & 94 / 100 (94\%) \\
Extraction confidence & 95 / 100 (95\%) \\
\bottomrule
\end{tabular}
\caption{Exact agreement of PSE outputs under repeated decoding for 100 randomly sampled ALM-generated captions.}
\label{tab:stab}
\end{table}

\subsection{PSE Capacity and Stability}

Replacing DeepSeek-32B with the smaller Qwen3-8B~\citep{yang2025qwen3technicalreport} achieves 68.2 (6.3)\% Macro-F1 on DAIC-WOZ, remaining above HAREN-CTC at 57.9 (2.1)\% and the Caption Embedding mean of 66.6 (14.2)\%.

Repeated decoding shows high exact agreement across all SC-Para dimensions (Table~\ref{tab:stab}). Sensitivity to meaning-preserving prompt paraphrases yields a mean evidence-score difference of 2.30 points, with correlations ranging from $.841$ to $.952$.

\subsection{Methodological Details for Statistical and Controlled Analyses}\label{app:analysis_details}

\paragraph{Semantic-Context Control:}
Because changing lexical content would also change the waveform, while
speech resynthesis could introduce acoustic artifacts, we isolate
semantic calibration at the PSE stage. The analysis uses 100 utterances
from 66 participants. For each utterance, we hold the ALM-generated
caption-level acoustic description fixed and compare three prompt
conditions: acoustic-only, non-depression-related context, and
depression-related context.

Depression-related context increases the PSE-derived evidence score by
8.6 points relative to non-depression-related context. The
utterance-paired comparison yields $t=8.64$, $p<.001$, and
$d_z=.86$; the corresponding participant-clustered 95\% confidence
interval is $[6.7,10.6]$. Depression-related context also increases the
score by 8.7 points relative to acoustic-only input, whereas
non-depression-related context differs from acoustic-only input by only
0.1 points ($p=.92$). This experiment characterizes semantic-context
sensitivity at the PSE stage while keeping the caption-level acoustic
description unchanged.

\paragraph{State Associations and Benjamini--Hochberg Correction:}
Figure~\ref{fig:interpretable} displays the ten highest-ranked state pairs for descriptive comparison. Seven pairs satisfy both prespecified screening criteria: $n\geq100$ utterances and a Benjamini--Hochberg-adjusted value of $q<.05$. The three lower-support configurations ($n=55$, $77$, and $92$) are shown only for context and are not interpreted. Because multiple utterances from each participant share one depression label, these utterance-level $q$ values do not account for within-participant clustering and must not be interpreted as participant-level confirmatory significance.

\section{Robustness, Bias, and Error Analysis} \label{app:robustness}

\subsection{Speaker-Identity and Additive Noise}
A speaker-verification probe on DAIC-WOZ (Table~\ref{tab:speaker}) yields ROC-AUC values close to chance for SC-Para (52.8--52.9\%), whereas WavLM reaches 78.7\%.

Mixing background noise from MUSAN~\citep{snyder2015musan} at 20, 10, and 5 dB SNR decreases Macro-F1 from 71.9\% to 68.6\%, 63.9\%, and 62.6\%, respectively.

\begin{table}[t]
\centering
\small

\resizebox{\columnwidth}{!}{
\begin{tabular}{lccc}
\toprule
Representation & Macro-F1 & AUC \\
\midrule
SC-Para (5 states) & 51.9 (0.8) & 52.8 (0.9) \\
SC-Para (7 dimensions) & 52.0 (0.8) & 52.9 (0.9) \\
WavLM (1024 dims) & 71.7 (0.3) & 78.7 (0.4) \\
\bottomrule
\end{tabular}
}
\caption{Speaker-verification probe on DAIC-WOZ.}
\label{tab:speaker}

\vspace{1em}

\resizebox{\columnwidth}{!}{
\begin{tabular}{lrr|lrr}
\toprule
\multicolumn{3}{c|}{DAIC-WOZ} & \multicolumn{3}{c}{MODMA} \\
Severity & $N$ & Acc (\%) & Severity & $N$ & Acc (\%) \\
\midrule
Minimal & 22 & 82.7 & Minimal & 26 & 92.3 \\
Mild & 11 & 54.5 & Mild & 4 & 100.0 \\
Mod. & 5 & 80.0 & Mod. & 3 & 93.3 \\
Mod-Sev. & 7 & 68.6 & Mod-Sev. & 10 & 80.0 \\
Severe & 2 & 100.0 & Severe & 9 & 95.6 \\
\bottomrule
\end{tabular}
}
\caption{Post-hoc prediction accuracy across PHQ severity strata for DAIC-WOZ official test participants and the complete MODMA out-of-fold cohort. Values are mean accuracies across the same five random seeds used in the main evaluation. $N$ denotes the number of unique participants in each severity stratum.
}
\label{tab:sev}

\vspace{1em}

\resizebox{\columnwidth}{!}{%
\begin{tabular}{lccc}
\toprule
Stage & Time/utt (s) & Total GPU (h) & Peak Mem (GiB) \\
\midrule
Captioner & 1.64 & 10.5 & 87.5 \\
PSE & 1.38 & 8.8 & 87.1 \\
ParaCalib (Total) & 3.0 & 19.3 & 87.5 \\
SSL baselines & 0.019--0.027 & 0.12--0.18 & 1.4--1.5 \\
\bottomrule
\end{tabular}
}
\caption{Representation-extraction cost on DAIC-WOZ.}
\label{tab:cost}

\end{table}

\subsection{Error Analysis and Computational Cost}
Consistent false positives lie closer to the PHQ-8 cutoff than reliable controls (mean PHQ-8 of 6.4 vs 2.6). Consistent false negatives contain fewer valid utterances on average than reliably classified true positives (66 vs.\ 119). Table~\ref{tab:sev} reports a post-hoc severity-stratified analysis. Table~\ref{tab:cost} summarizes computational costs. Table~\ref{tab:discard} reports discard rates due to \texttt{unknown} assignments, which do not substantially alter class composition.

\begin{table}[t]
\centering
\small

\resizebox{\columnwidth}{!}{%
\setlength{\tabcolsep}{6pt}
\begin{tabular}{llccccc}
\toprule
Dataset & Split & Overall & \multicolumn{2}{c}{Utterance discard rate} & \multicolumn{2}{c}{Per-participant discard rate} \\
\cmidrule(lr){4-5} \cmidrule(lr){6-7}
& & & Non-dep. & Dep. & Non-dep. & Dep. \\
\midrule
\multirow{3}{*}{DAIC-WOZ}
& Train & 5.7\% & 6.0\% & 4.9\% & 6.1 (3.8)\% & 4.4 (3.5)\% \\
& Dev & 4.7\% & 4.6\% & 4.8\% & 5.0 (3.0)\% & 4.5 (2.2)\% \\
& Test & 6.2\% & 6.2\% & 6.2\% & 6.0 (3.7)\% & 7.0 (3.0)\% \\
\midrule
\multirow{3}{*}{MODMA}
& Train & 1.9\% & 1.4\% & 2.6\% & 1.4 (3.9)\% & 2.6 (4.7)\% \\
& Dev & 2.5\% & 2.8\% & 2.2\% & 2.8 (2.8)\% & 2.3 (2.8)\% \\
& Test &0.6\% & 0.0\% & 1.1\% & 0.0 (0.0)\% & 1.1 (2.2)\%\\
\bottomrule
\end{tabular}
}
\caption{Discard rates after PSE filtering by split and label group.}
\label{tab:discard}

\vspace{1em}

\resizebox{\columnwidth}{!}{%
\begin{tabular}{lccccc}
\toprule
\multirow{2}{*}{Validation Target} & \multicolumn{3}{c}{Consistency Metrics} & \multicolumn{2}{c}{Error Counts} \\
\cmidrule(lr){2-4} \cmidrule(lr){5-6}
& Pos. Judgments & Strict Pass & Rater Agree. & Inconsist. by $\ge1$ & Inconsist. by both \\
\midrule
Caption consistency & 38/40 (95\%) & 18/20 (90\%) & 18/20 (90\%) & 2/20 & 0/20 \\
PSE constituent states & 36/40 (90\%) & 17/20 (85\%) & 18/20 (90\%) & 3/20 & 1/20 \\
\bottomrule
\end{tabular}
}
\caption{Human validation of generated captions and SC-Para states on 20 sampled utterances.}
\label{tab:human_val}

\end{table}

\subsection{Attention-Weighted Dispersion Measure}
\label{app:dispersion}

To assess the internal consistency of participant-level representations, we computed an attention-weighted dispersion measure across valid utterance-level SC-Para vectors. Lower values indicate that the utterances receiving higher attention weights have more similar categorical SC-Para patterns within a participant. Group differences were evaluated using a subject-level permutation test, with a bootstrap confidence interval for the mean difference, a Mann--Whitney $U$ test, and Cliff's delta as an effect-size estimate.

In the DAIC-WOZ test split, depression-label participants had lower mean attention-weighted dispersion than non-depression-label participants (0.3617 vs.\ 0.3945; $n=14$ and $n=33$, respectively). The mean difference was $-0.0327$ (bootstrap 95\% CI $[-0.0593,-0.0081]$; subject-level permutation $p=.0189$). The result was consistent with the Mann--Whitney $U$ test ($p=.0467$), with Cliff's $\delta=-0.3723$. Given the small test-set groups, this analysis is considered exploratory.

\subsection{Human Validation}\label{app:human_validation}
Two annotators (graduate researchers proficient in the evaluated languages) evaluated 20 randomly sampled utterances. Since this small-scale validation was conducted internally as part of routine research activities, no external recruitment or additional financial compensation was involved. The generated captions and the five PSE-derived constituent states were judged consistent with perceived vocal evidence in 18/20 and 17/20 samples, respectively (Table~\ref{tab:human_val}).

\section{Prompts and Output Schema} \label{app:prompts}
Table~\ref{tab:prompts_combined} presents the content-masking prompt, an abbreviated SC-Para feature schema, and the PSE extraction template used in ParaCalib.

\begin{table*}[t]
\centering
\small
\begin{tabular}{|p{0.96\textwidth}|}
\hline
\rowcolor{gray!15} \textbf{Prompt 1: Content Masking (Lexical Cue Removal)} \\
\hline
\begin{minipage}{0.94\textwidth}
\begin{Verbatim}[breaklines=true,breakanywhere=true,fontsize=\footnotesize]
You are an audio description filtering expert. Your task is to filter out speech content transcription from audio descriptions and keep only the audio feature descriptions.
Content to REMOVE:
1. Specific words and phrases spoken by the speaker (direct quotes)
2. Detailed descriptions of conversation content
3. Explanations of semantic meaning and topics discussed

Content to KEEP:
1. Speaker's voice characteristics (timbre, tone, pitch, voice quality)
2. Speaking patterns (pace, rhythm, pauses, cadence)
...
Return ONLY the filtered description without any explanations or additional comments. If the entire description is transcription content, return an empty string.
\end{Verbatim}
\end{minipage} \\

\hline
\rowcolor{gray!15} \textbf{Prompt 2: SC-Para Feature Definitions \& Schema} \\
\hline
\begin{minipage}{0.94\textwidth}
\begin{Verbatim}[breaklines=true,breakanywhere=true,fontsize=\footnotesize]
Feature definitions:
prosody.pitch_variability:
   - "flat": reduced pitch variation, monotonous or restricted intonation.
   - "moderate": normal or balanced pitch variation after semantic calibration.
   - "high": large pitch variation that cannot be explained only by transcript context.
   - "unknown": insufficient evidence.
... [Other definitions omitted for brevity: speaking_rate, rhythm, pauses, energy] ...

depression_related_paralinguistic_evidence.value:
   - A calibrated depression-related paralinguistic evidence score from 0 to 100.
depression_related_paralinguistic_evidence.confidence:
   - A value from 0.0 to 1.0. Higher confidence means the evidence is clear.
\end{Verbatim}
\end{minipage} \\

\hline
\rowcolor{gray!15} \textbf{Prompt 3: PSE Extraction Template} \\
\hline
\begin{minipage}{0.94\textwidth}
\begin{Verbatim}[breaklines=true,breakanywhere=true,fontsize=\footnotesize]
You are an expert in paralinguistic analysis of depression-related speech patterns.
IMPORTANT CONSTRAINTS:
- You CANNOT infer identity, mental diagnosis, or life situation.
- You are NOT making a medical diagnosis.
- Your task is to estimate the strength of depression-related paralinguistic evidence at the utterance level ONLY.

Input:
A caption generated by an audio-language model for a single utterance.

Your task:
1. Extract acoustically observable vocal and paralinguistic information.
2. Based on vocal and paralinguistic patterns, estimate the presence and strength of depression-related speech characteristics.
3. If vocal information is sparse or ambiguous, use "unknown" and reduce confidence.

Output STRICTLY in the following JSON format. Each JSON corresponds to ONE utterance.
{
  "prosody": {
    "pitch_variability": "flat/moderate/high/unknown",
    ...
  },
  "depression_related_paralinguistic_evidence": {
    "value": 0-100,
    "confidence": 0.0-1.0
  }
}
\end{Verbatim}
\end{minipage} \\
\hline
\end{tabular}
\caption{Selected prompts and an abbreviated output schema used for content masking and PSE extraction.}
\label{tab:prompts_combined}
\end{table*}
\end{document}